\documentclass{mnras}
\usepackage{graphicx}
\usepackage{times}
\usepackage[dvipsnames]{xcolor}
\usepackage{color}

\catcode`\@=11
\newcommand\gsim{\ifmmode{\mathrel{\mathpalette\@versim>}}
    \else{$\mathrel{\mathpalette\@versim>}$}\fi}
\newcommand\lsim{\ifmmode{\mathrel{\mathpalette\@versim<}}
    \else{$\mathrel{\mathpalette\@versim<}$}\fi}
\catcode`\@=12

\newcommand\rhos{\rho_*}
\newcommand\rhog{\rho_{\rm g}}
\newcommand\rhogamma{\rho_{\gamma}}
\newcommand\rhoDM{\rho_{\rm DM}}
\newcommand\rhon{\rho_{\rm n}}
\newcommand\rhoJ{\rho_{\rm J}}
\newcommand\rhoNFW{\rho_{\rm NFW}}

\newcommand\rs{r_*}
\newcommand\rg{r_{\rm g}}
\newcommand\rgamma{r_{\gamma}}
\newcommand\rh{r_{\rm h}}
\newcommand\rj{r_{\rm J}}
\newcommand\reff{R_{\rm e}}
\newcommand\ra{r_{\rm a}}
\newcommand\sa{s_{\rm a}}
\newcommand\rt{r_{\rm t}}

\newcommand\Phig{\Phi_{\rm g}}
\newcommand\Phibh{\Phi_{\rm BH}}
\newcommand\Phidm{\Phi_{\rm DM}}
\newcommand\PhiJ{\Phi_{\rm J}}
\newcommand\Phis{\Phi_*}
\newcommand\PsiT{\Psi_{\rm T}}
\newcommand\Psig{\Psi_{\rm g}}
\newcommand\Psin{\Psi_{\rm n}}

\newcommand\Ms{M_*}
\newcommand\Mj{M_{\rm J}}
\newcommand\Mbh{M_{\rm {BH}}}
\newcommand\Mg{M_{\rm g}}
\newcommand\MT{M_{\rm T}}
\newcommand\Mgamma{M_{\gamma}}
\newcommand\Mp{M_{\rm p}}
\newcommand\MpJ{M_{\rm pJ}}
\newcommand\MR{{\cal R}}
\newcommand\Rm{{\mathcal{R}}_{\rm m}}
\newcommand\Rmon{{\mathcal{R}}_{\rm mon}}
\newcommand\Rcmz{{\mathcal{R}}_{\rm CMZ}}
\newcommand\Rdm{{\mathcal{R}}_{\rm {DM}}}

\newcommand\qt{\tilde q}
\newcommand\Fp{F_{+}}
\newcommand\Fm{F_{-}}
\newcommand\Er{\mathcal{E}}

\newcommand\It{\mathcal{I}_{\rm g}}
\newcommand\At{\mathcal{A}_{\rm g}}
\newcommand\Ibh{\mathcal{I}_{\rm BH}}
\newcommand\Abh{\mathcal{A}_{\rm BH}}

\newcommand\srad{\sigma_{\rm r}}
\newcommand\stan{\sigma_{\rm t}}

\newcommand\Ks{K_*}
\newcommand\Krad{K_{\rm *r}}
\newcommand\Ktan{K_{\rm *t}}
\newcommand\Ws{W_*}
\newcommand\Wss{W_{**}}
\newcommand\Wg{W_{\rm *g}}
\newcommand\Wbh{W_{\rm *BH}}
\newcommand\Wdm{W_{\rm *DM}}
\newcommand\Us{U_*}
\newcommand\Uss{U_{**}}
\newcommand\Ug{U_{\rm *g}}
\newcommand\Ubh{U_{\rm *BH}}
\newcommand\Udm{U_{\rm *DM}}
\newcommand\Bg{B_{*\rm g}}

\newcommand\vc{v_{0}}
\newcommand\Sigs{\Sigma_{\gamma}}
\newcommand\SigJ{\Sigma_{\rm J}}

\newcommand\sigp{\sigma_{\rm p}}

\newcommand\sam{\sa^{-}}
\newcommand\sap{\sa^{+}}
\newcommand\cond{{\cal C}}
\newcommand\condm{\cond_{-}}
\newcommand\condp{\cond_{+}}
\newcommand\MD{M_{\rm {DM}}}

\newcommand\csih{\xi_{\rm NFW}}

\title[Two--component galaxy models with central BH]{Two-component Jaffe models with a
  central black hole. I: the spherical case}

   \author[L. Ciotti, A. Ziaee~Lorzad]
          {Luca Ciotti $^1$, Azadeh Ziaee~Lorzad$^1$
\\$^1$Department of Physics and Astronomy, 
      University of Bologna, via Gobetti 93/3, 40129 Bologna, Italy}

\date{Submitted, July 21, 2017 - Resubmitted, October 14, 2017}
\pubyear{...}

\begin{document} 
\maketitle

\begin{abstract} 

  Dynamical properties of spherically symmetric galaxy models where
  both the stellar and total mass density distributions are described
  by the Jaffe (1983) profile (with different scale-lenghts and
  masses), are presented.  The orbital structure of the stellar
  component is described by Osipkov--Merritt anisotropy, and a black
  hole (BH) is added at the center of the galaxy; the dark matter halo
  is isotropic.  First, the conditions required to have a nowhere
  negative and monothonically decreasing dark matter halo density
  profile, are derived. We then show that the phase-space distribution
  function can be recovered by using the Lambert-Euler $W$ function,
  while in absence of the central BH only elementary functions appears
  in the integrand of the inversion formula. The minimum value of the
  anisotropy radius for consistency is derived in terms of the galaxy
  parameters.  The Jeans equations for the stellar component are
  solved analytically, and the projected velocity dispersion at the
  center and at large radii are also obtained analytically for generic
  values of the anisotropy radius. Finally, the relevant global
  quantities entering the Virial Theorem are computed analytically,
  and the fiducial anisotropy limit required to prevent the onset of
  Radial Orbit Instability is determined as a function of the galaxy
  parameters. The presented models, even though highly idealized,
  represent a substantial generalization of the models presentd in
  Ciotti et al. (2009), and can be useful as starting point for more
  advanced modeling the dynamics and the mass distribution of
  elliptical galaxies.

\end{abstract}


\section{Introduction}

Spherically symmetric galaxy models, despite their simplicity, are
useful tools for theoretical and observational works in Stellar
Dynamics, and for the modelization of stellar systems (e.g., Bertin
2000, Binney \& Tremaine 2008). Quite obviously spherical symmetry is
an oversimplification when considering the vast majority of stellar
systems, and a useful spherical model must compensate this limitation
with other features, that make its use preferred or even recommended,
especially in preliminary investigations. Among the important features
of a useful spherical model here we list analytical simplicity,
structural and dynamical flexibility, i.e., possibility to add to the
stellar component a dark matter halo with adjustable density profile,
or alternatively to specify the total density profile, to include the
dynamical effects of a central black hole, to control orbital
anisotropy

For example, the density profile of the stellar distribution of the
model, once projected, should be similar to that of early-type
galaxies, i.e. to the de Vaucouleurs (1948) $R^{1/4}$ law, or better,
to its generalization, the so-called $R^{1/m}$ law (Sersic 1963).
Unfortunately the $R^{1/m}$ law doesn't allow for an explicit
deprojecton in terms of elementary functions, however the so-called
$\gamma$ models (Dehnen 1993, Tremaine et al. 1994) in projection are
well fitted over a large radial range, by the $R^{1/m}$ law. This is
especially true for the Jaffe (1983) and Hernquist (1990) models.

Another important feature of a useful spherical model is the
possibility to reproduce the large scale observational properties of
the {\it total} density profile of early-type galaxies.  In fact,
analysis of stellar kinematics (e.g. Bertin et al. 1994, Rix et
al. 1997, Gerhard et al. 2001), as well as several studies combining
stellar dynamics and gravitational lensing support the idea that the
dark and the stellar matter in elliptical galaxies are distributed so
that their total mass profile is described by a density distribution
proportional to $r^{-2}$ (e.g., see Treu \& Koopmans 2002, 2004; Rusin
et al. 2003; Rusin \& Kochanek 2005; Koopmans et al. 2006; Gavazzi et
al. 2007; Czoske et al. 2008; Dye et al. 2008, Nipoti et al. 2008, see
also Shankar et al. 2017). It is clear that simple dynamical models of
two-component galaxies can be useful as starting point of more
sophisticated investigations based on axysimmetric or triaxial galaxy
models (e.g., Cappellari et al. 2007, van den Bosch et
al. 2008). Simple models with flat rotation curve have been in fact
constructed (e.g. Kochaneck 1994, Naab \& Ostriker 2007). In
particular we recall the family of two-component galaxy models whose
total mass density is proportional to $r^{-2}$, while the visible
(stellar) mass is described by the $\gamma$ models (Ciotti et
al. 2009, hereafter CMZ09; see also the double power-law models of
Hiotelis 1994). These latter models have been used in
hydrodynamical simulations of accretion onto the central supermassive
black hole (hereafter, BH) in elliptical galaxies (Ciotti \& Ostriker
2012, and references therein).  We notice that other models built with
the same approach have been recently applied for the interpretation of
observations (Poci et al. 2017).  We also remark that the approach
used to build these models is different from the standard one, where a
dark matter halo (herefater, DM) is {\it added} to the stellar
distribution (e.g. Ciotti \& Renzini 1993; Ciotti et al. 1996,
hereafter CRL96; Ciotti 1996, 1999; Sect. 4.4 in CMZ09 ).

A third important feature of a useful spherical model, strictly
related to the previous point, is the possibility to easily compute
the dynamical properties of the stellar component in presence of a
central BH, and possibly to be proved dynamically consistent (see
Sect. 3.1). In fact, supermassive BHs with a mass of the order of
$\Mbh\simeq 10^{-3}\Ms$ are routinely found at the center of the
stellar sferoids of total mass $\Ms$ (e.g., see Magorrian et al. 1988,
Kormendy \& Ho 2013).

Following the arguments above, this paper builds on the CMZ09 model,
and present an even more general (and realistic) class of models,
containing the CMZ09 model as a limit case. On one side, we maintain
the assumption of a Jaffe profile for the stellar distribution, but
now the total density profile is described by another Jaffe law
(instead of a pure $r^{-2}$ law), so that the total mass of the models
(that we call JJ models) is finite. At the same time, the scale-lenght
of the total density is a free parameter and so we can reproduce an
$r^{-2}$ profile over an arbitrary large radial range. Finally, a
central BH of arbitrary mass (missing in CMZ09 models) is considered
when solving the dynamical equations. For JJ models we show that the
Jeans equations for the stellar component with Osipkov-Merritt
(Osipkov 1979, Merritt 1985a, hereafter OM) radial anisotropy can be
solved analytically, and the projected velocity dispersion at the
center and at large radii can be expressed by means of extremely
simple formulae for generic values of the model parameters. The
positivity of the phase-space density distribution function of the
stellar component (hereafter DF), the so-called consistency, is easily
investigated by using a remarkable property of JJ models, i.e. the
fact that the radial coordinate can be written in terms of the total
potential in terms of the so-called Lambert-Euler $W$ function. By
using this property, we determine the maximum amount of radial
anisotropy allowable for consistency as a function of the galaxy
parameters. These results add to the large amount of phase-space
information already available about one and two-component $\gamma$
models (e.g., Carollo et al. 1995, Ciotti 1996, 1999; Baes et
al. 2005, Buyle et al. 2007, Ciotti \& Morganti 2009).  As a byproduct
of our analysis we also found that the one-component Jaffe model, at
variance with statements in the literature, {\it cannot} be supported
by purely radial orbits.  We note that the $W$ function also appears
in the recenty discovered analytical solution of the isothermal Bondi
accretion problem in Jaffe galaxies with central BH (Ciotti \&
Pellegrini 2017), and this fact suggests a first natural application
of JJ models outside the field of Stellar Dynamics, namely in the
field of BH accretion and AGN feedback.

The paper is organized as follows. In Section 2 the main structural
properties of the models are presented. In Section 3 an investigation
of the phase-space properties of the models is carried out both from
the point of view of necessary and sufficient conditions for
consistency, and from direct inspection of the DF. In Section 4 the
solution of the Jeans equation with OM radial anisotropy is presented,
together with the projection of the velocity dispersion profile at
small and large radii.  In Section 5 the important properties related
to the Virial Theorem and global energetic are explicitly calculated,
and the maximum amount of radial anisotropy that can be sustained by
the model without developing Radial Orbit Instability is
estimated. The main results are summarized in Section 6, while more
technical details are given in the Appendix.

\section{THE MODELS}

As anticipated in the Introduction, the present models are
characterized by a {\it total} density distribution (stars plus DM) $\rhog$ described by a
Jaffe (1983) profile; the stellar density distribution $\rhos$ is
also described by a Jaffe profile, in general with a different scale
radius. For future use we recall that the Jaffe density of total mass 
$\Mj$ and scale length $\rj$ is given by
\begin{equation}
\rhoJ(r)={\Mj \rj\over 4\pi r^2(\rj + r)^2}.
\end{equation}
The
cumulative mass contained within the sphere of radius $r$,
and the associated gravitational potential (with the natural condition
of vanishing at infinity, pertintent to systems of finite mass), are
given by 
\begin{equation}
\Mj(r)={\Mj r\over \rj+r},\quad
\PhiJ(r)={G\Mj\over\rj}\ln {r\over \rj+r}.
\end{equation}

Moreover the Jaffe model belongs to the family of $\gamma$-models
\begin{equation}
\rhogamma (r)={(3-\gamma)\Mgamma \rgamma \over 4\pi
  r^{\gamma}(\rgamma+r)^{4-\gamma}},\quad
0\leq\gamma <3, 
\end{equation}
where $\Mgamma$ is the total mass, $\rgamma$ is a scale-length, and
eq. (1) is obtained for $\gamma=2$. The cumulative mass within the
sphere of radius $r$ is given by
\begin{equation}
\Mgamma (r)=\Mgamma\times\left({r\over\rgamma +r}\right)^{3-\gamma},
\end{equation}
so that the half-mass (spatial) radius is $\rh =\rgamma/(2^{1\over
  3-\gamma}-1)$, and $\rh=\rj$ for the Jaffe model.  For generic
values of $\gamma$ the projected density at radius $R$ in the
projection plane is given by
\begin{equation}
\Sigs(R)=2\int_R^{\infty}{\rhogamma (r)rdr\over\sqrt{r^2-R^2}},
\end{equation}
(e.g., Binney \& Tremaine 2008), but unfortunately it cannot be
expressed in terms of elementary functions.  However, for $\gamma=2$
\begin{equation}
\SigJ (R)={\Mj\over\rj^2}\times
\cases{
            \displaystyle{
{1\over 4\eta} +
{\sqrt{1-\eta^2}-(2-\eta^2) {\rm arcsech(\eta)}\over 2\pi (1-\eta^2)^{3/2}}},
                                                     \quad 0<\eta <1 ;
         \cr\cr
         \displaystyle{{1\over 4} - {2\over 3\pi}},
                       \quad\quad\quad\quad\quad\quad\quad \eta=1;
         \cr\cr
         \displaystyle{
{1\over 4\eta}-
{\sqrt{\eta^2-1}+(\eta^2-2){\rm arcsec(\eta)}\over 2\pi(\eta^2-1)^{3/2}}},
                      \quad \eta >1;}
\end{equation}
where $\eta\equiv R/\rj$. In the central and in the very external
regions the projected density profile behave like a power law, with
\begin{equation}
\SigJ (R)\sim {\Mj\over\rj^2} \times\cases{
         \displaystyle{{1\over 4\eta}},
                                                     \quad R\to 0;
         \cr\cr
         \displaystyle{{1\over 8\eta^3}},
                       \quad R\to \infty,
         \cr\cr
         }
\end{equation}
respectively.  Finally, an important structural property that we will
consider in the following is the projected mass $\Mp(R)$ contained
within the cylinder of radius $R$. It can be proved that for spherical
systems of finite total mass
\begin{equation}
\Mp(R)\equiv2\pi\int_0^R \Sigma (R)\, R\, dR =M-4\pi \int_R^\infty \rho(r)r\sqrt{r^2-R^2}dr.
\end{equation}
It follows that the projected mass of the Jaffe model is given by $\MpJ (R)=\Mj \times g(\eta)$, where 
\begin{equation}
g(\eta)=\eta \times \cases{
            \displaystyle{{\pi\over 2}-{\eta\, {\rm arcsech}(\eta)\over \sqrt{1-\eta^2}}},
                                                     \quad 0<\eta <1 ;
         \cr\cr
         \displaystyle{{\pi\over 2}-1},
                       \quad\quad\quad\quad\quad\quad\quad \eta=1;
         \cr\cr
         \displaystyle{{\pi\over 2}-{\eta \, {\rm arcsec} (\eta)\over \sqrt{\eta^2-1}}},
                      \quad \eta >1.}
\end{equation}
In particular, the effective radius $\reff$ of the Jaffe profile
(i.e., the radius in the projection plane encircling half of the total
mass), where $g(\eta_{\rm e})=1/2$, is $\reff\simeq 0.7447\rj$ (in the
Jaffe original paper the slightly erroneous value of 0.763 is
reported).

\subsection{Stellar and total mass distribution}

We denote our family of models as ``JJ'' models, to indicate that it
is a two-component Jaffe model, even though constructed in a different
way with respect to other two-components Jaffe models in the
literature (CLR96, Ciotti 1996, 1999). The properties of the stellar
component are obtained with $\Mj=\Ms$ and $\rj=\rs$ in eqs. (1)-(9),
while for the galaxy total density distribution (stars plus DM)
$\Mj=\Mg$ and $\rj=\rg$.  We adopt $\Ms$ and $\rs$ as the natural mass
and length scales, and we define
\begin{equation}
 s\equiv {r\over\rs}, \quad \xi \equiv {\rg\over\rs},\quad \MR\equiv
 {\Mg\over\Ms} =\Rdm +1.
\end{equation}
From the request that the DM component has a non-negative total mass
$\MD$ it follows that $\Rdm\equiv\MD/\Ms\geq 0$, and so $\MR\geq
1$. It is important to note that the request of a non-negative $\MD$
does not prevent the possibility of an unphysical, {\it locally negative} DM
density. This case will be excluded with the introduction of an
additional constraint, determined in Sect. 2.2.
We also define
\begin{equation}
\rhon\equiv{\Ms\over 4\pi\rs^3},\quad
\Psin\equiv{G\Ms\over\rs},
\end{equation}
as the natural density and potential scales.
With these conventions,
eqs. (1) and (2) for the galaxy model become
\begin{equation}
\rhog(r)= {\MR\xi\rhon\over s^2(\xi+s)^2},
\end{equation}
and
\begin{equation}
\Mg(r)={\Ms \MR s\over \xi+s},\quad
\Phig (r)={\MR\Psin\over\xi}\ln {s\over \xi+s}.
\end{equation}
We note here an important connection of the JJ models with the models
in CMZ09. In fact, the total galaxy density profile in CMZ09
(eq. [6] therein) can be written as
\begin{equation}
\rhog^{\rm{CMZ}}(r)={\vc^2\over 4\pi G r^2}={\Rcmz\rhon\over s^2},
\end{equation}
where $\vc$ is the constant circular velocity. As the total mass
associated with eq. (14) diverges, the parameter $\Rcmz =\vc^2/\Psin$
is {\it not} the ratio of the total-to-stellar mass as in JJ
models. An elementary integration shows that $\Rcmz=\Mg(\rs)/\Ms$,
i.e., it is the {\it total} mass contained within the half mass radius
of the Jaffe stellar density profile, normalized to the total stellar
mass.  From eqs. (12) and (14) it follows that the total density
distribution (and the associated quantities, such as the cumulative
and the projected mass profiles, and the force field) of CMZ09 models
can be obtained from JJ models with the substitution
\begin{equation}
\MR =\Rcmz\xi
\end{equation}
in the corresponding quantities, and then considering the limit for
$\xi\to \infty$. Some care is needed for the case of the
potential. In fact JJ models have finite mass and vanishing potential
at infinity, while the logarithimic potential of CMZ09 models
\begin{equation}
\Phig^{\rm{CMZ}}(r) =\vc^2\ln s,
\end{equation}
diverges for $s\to \infty$.  The proper way to reobtain
$\Phig^{\rm{CMZ}}$ from eq. (13) is to apply the substitution (15) to
JJ scaled potential $\Phig + \MR\Psin (\ln \xi)/\xi$, and then to take
the limit $\xi \to \infty$.

\subsection{The dark matter distribution: positivity and monotonicity}

Before studying the dynamical properties of the models, it is
important to determine the conditions for the positivity and radial
monotonicity of the density distribution of the DM halo. While as
anticipated in Sect. 2.1 the request of positivity is natural, a brief
comment is in order to justify the requirement of monotonicity. In fact,
it can be shown that monotonicity of the density as a function of the
potential is necessary for the positivity of the phase-space
distribution function. From the second Newton's Theorem, the
gravitational potential of a spherical system is necessarily radially
monotone, so that the density profile must be a monotone function of
radius (Ciotti \& Pellegrini 1992, hereafter CP92, see also Sect. 3).

As already found in the simpler two component model of CMZ09, also in
JJ models not all values of $\MR$ and $\xi$ are compatible with a
nowhere negative DM distribution $\rhoDM$. Curiously, it is
possible to obtain analytically the positivity condition for the more
general family of two-component $\gamma$ models, built with the same
approach of JJ models. For $\gamma\gamma$ models the DM
distribution can be written as:
\begin{equation}
\rhoDM(r)={(3-\gamma)\rhon\over s^{\gamma} }
\left[{\MR\xi\over (\xi+s)^{4-\gamma}} - {1\over (1+s)^{4-\gamma}} \right]:
\end{equation}
note that $\rhoDM$ of $\gamma\gamma$ models {\it is not a $\gamma$
  model, unless the stellar and total length scale are equal}, so that
in general the local DM-to-stellar mass ratio $\rhoDM (r)/\rhos (r)$
depends on $r$. It is easy to verify that the total DM mass
  associated with $\rhoDM$ is $\MD =4\pi\rhon\rs^3 (\MR -1)$.

\begin{figure*}
\includegraphics[height=0.28\textheight,width=0.33\textwidth]{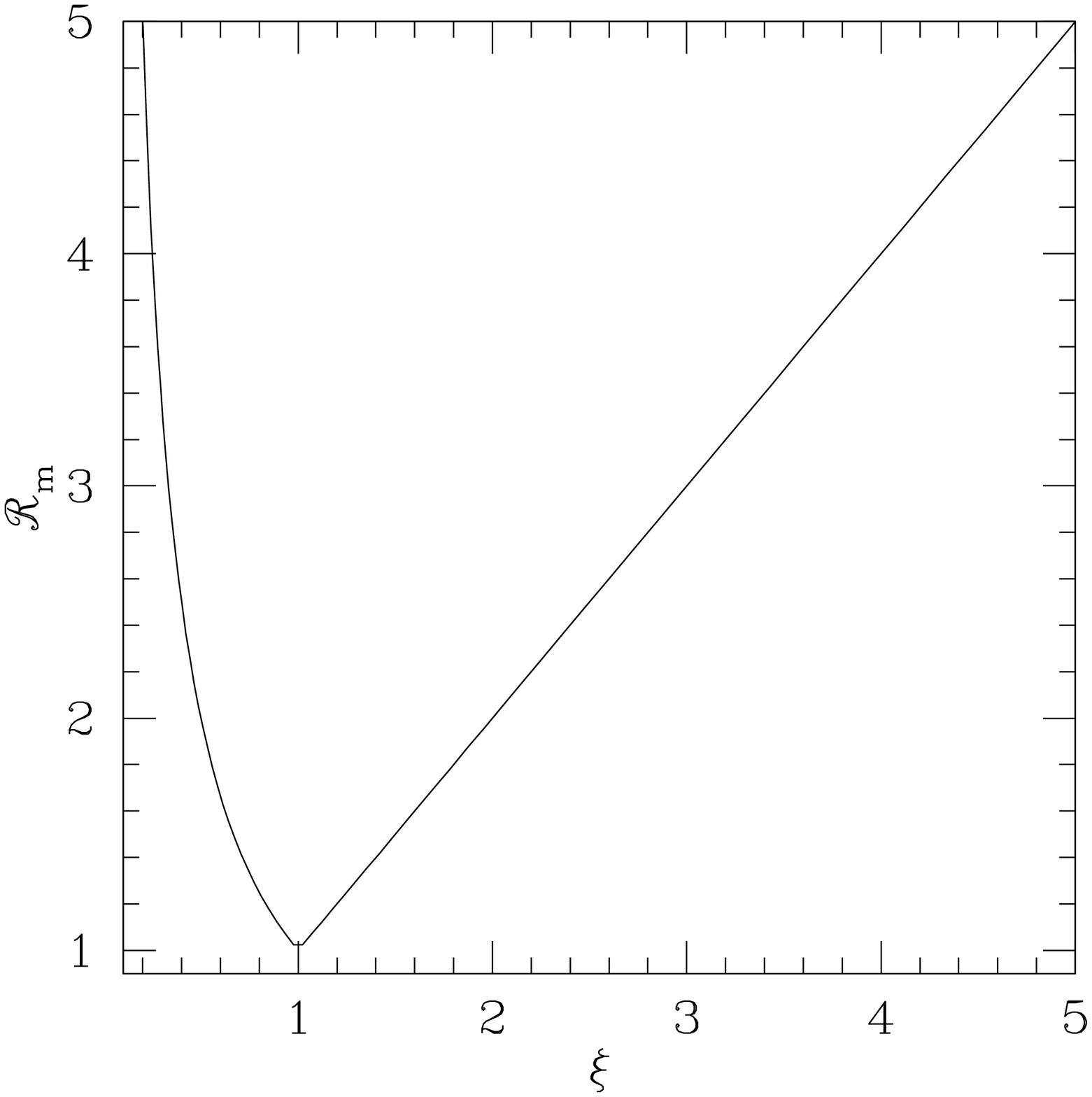}
\includegraphics[height=0.28\textheight,width=0.33\textwidth]{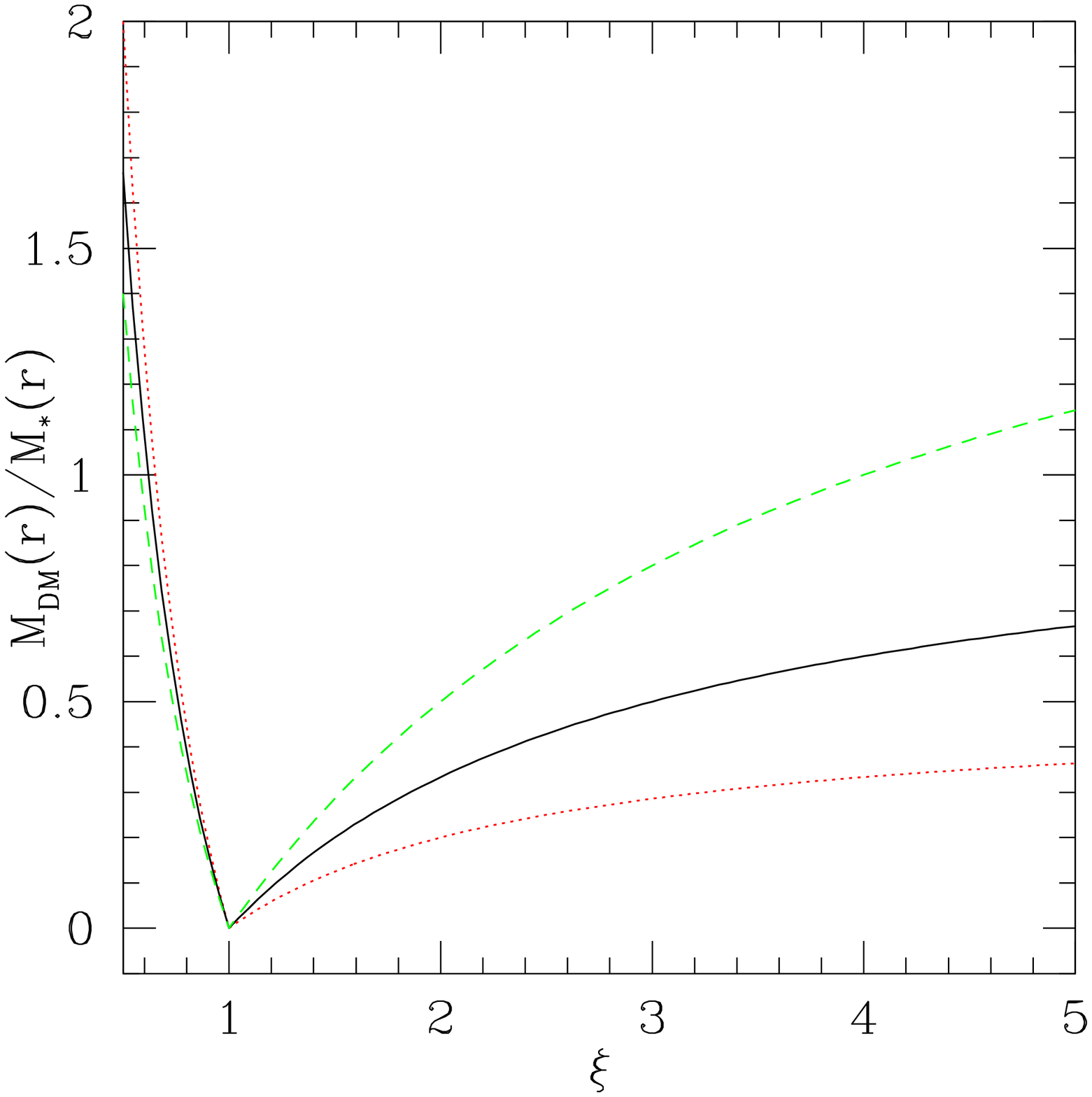}
\includegraphics[height=0.28\textheight,width=0.33\textwidth]{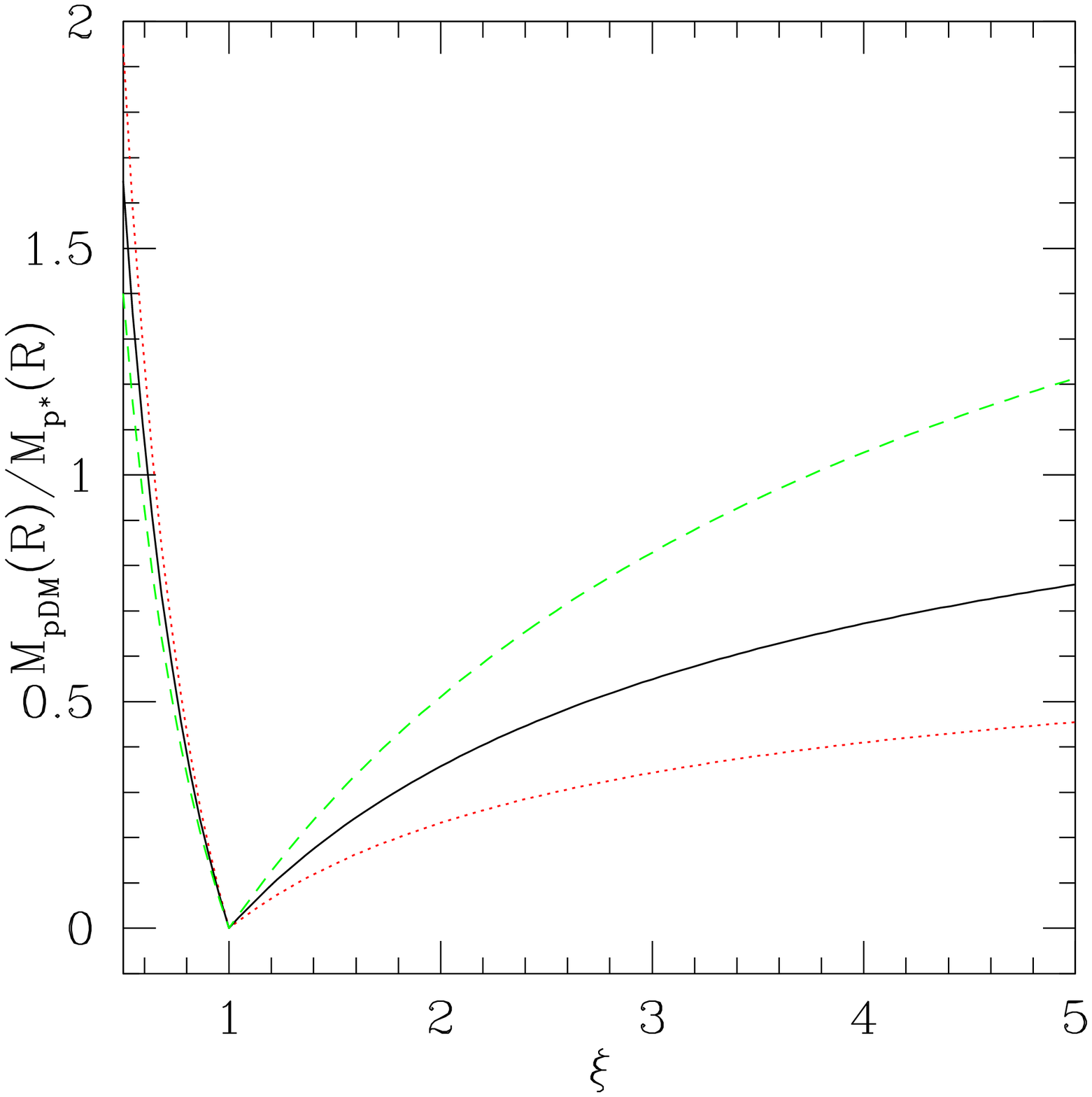}
\caption{Left panel: the minimum value of the total-to-stellar mass
  ratio $\Rm$, as a function of $\xi=\rg/\rs$, as given by
  eq. (18). Only models in the open triangular region are
  characterized by a DM halo with a nowhere negative density
  $\rhoDM$. Central panel: the minimum value of the volumic
  DM-to-stellar mass ratio inside a sphere of radius $r=0.5 \rs$,
  $\rs$, and $2\rs$ (red dotted, black solid, green dashed lines,
  respectively), as a function of $\xi$. Right panel: the minimum value
  of the projected DM-to-stellar mass ratio inside the circle of radius
  $R=0.5\reff$, $\reff$, and $2\reff$ (red dotted, black solid, green
  dashed lines, respectively), as a function of $\xi$.}
\label{fig:RminCsi}
\end{figure*}

In Appendix A we determine, for given $ 0\le \gamma <3$, the
conditions on $\MR$ and $\xi$ to have $\rhoDM \ge 0$ for $r\ge 0$. In the
case of the JJ models (i.e. $\gamma\gamma$ models with $\gamma=2$),
the positivity condition (A2) reduces to
\begin{equation}
\MR \ge\Rm(\xi)= \max\left({1\over \xi},\xi \right);
\end{equation}
a DM halo of a model with $\MR=\Rm$ is called a {\it minimum
  halo}. From equation above it follows that more and more DM is
needed for a total density distribution $\rhog$ more and more
contracted, or more and more expanded than the stellar distribution
$\rhos$. The minimum value $\Rm=1$ can be only adopted when $\xi=1$,
i.e. when the stellar and total density are proportional, and so
$\rhoDM$ can vanish everywhere. The situation is illustrated in Fig. 1
(left panel).

As anticipated the positivity of $\rhoDM$ is just a first condition
for the viability of the model.  A second request is the monotonicity
of $\rhoDM$ as a function of radius, and this reduces to the
determination of the minimum value $\Rmon$ so that $d\rhoDM /dr \leq
0$. The explicit discussion of this additional restriction is given in
Appendix A, for the whole family of $\gamma\gamma$ models. In
particular we found that for $1\leq\gamma<3$ (the range containing JJ
models, or the analogous two-component Hernquist models), the
positivity and monotonicity conditions for $\rhoDM$ coincide,
i.e. $\Rmon (\xi) =\Rm(\xi)$.

Equation (18) allows to discuss the relative trend of DM and stars in
JJ models, both at large radii and near the center, as a function of
$\MR$ and $\xi$.  For $r\to\infty$ and $\xi >1$ it is easy to show
that $\rhoDM \sim (\MR \xi -1)\rhos$, and so in the outskirts DM and
stars are proportional. When $\xi <1$ instead the situation is more
complicated: while in non minimum halo models $\rhoDM\sim (\MR/\xi
-1)\rhos$ and so DM and stars distributions are again proportional, in
the minimum halo case $\rhoDM \sim 2(1-\xi)\rhos/s\propto r^{-5}$, so
that the galaxy is baryon-dominated in the external regions. The
situation inverts for $r\to 0$. In fact, in this case for $\xi <1$ we
have $\rhoDM \sim (\MR/\xi -1)\rhos$ so that DM and stars mass are
locally proportional, but for $\xi >1$, while in non minimum halo
models $\rhoDM \sim (\MR/\xi-1)\rhos$, in the minimum-halo models
$\rhoDM \sim 2(1-1/\xi)\rhos s\propto r^{-1}$, so that these models
are centrally baryon-dominated.

It can be of interest for applications to evaluate the relative amount
of dark and visible mass within a prescribed (spatial or projected)
radius. The minimum value for this quantity is easily calculated from
eqs. (2) and (13),
\begin{equation}
{\MD(r)\over\Ms(r)}\geq \frac{\Rm(\xi)(1+s)}{\xi+s}-1,
\end{equation}
where $\MD(r)=\Mg(r)-\Ms(r)$. In Fig. 1 (middle panel) the mass ratios
corresponding to three representative values of $r$ are shown as a
function of $\xi$. For example in the case of a sphere of radius equal
to a half mass radius of the stellar distribution (i.e. $r=\rs$),
the minimum value $\MD/\Ms$ is less than unity for $\xi >1$:
this is a significant improvement of JJ models with respect to the models
of CMZ09, where this ratio can not be less than unity (see Fig.3
therein).

A similar behavior is obtained for the ratio of projected
DM-to-visible mass within some prescribed aperture $R$, and from the
eq. (9) it is easy to show that
\begin{equation}
{M_{\rm{pDM}}(R)\over M_{\rm p*}(R)} \ge {\Rm(\xi) g(\eta/\xi)\over g(\eta)}-1.
\end{equation}
In Fig. 1 (right panel) we plot this quantity as a function of $\xi$
for three representative values of the aperture radius, i.e. $\reff/2,
\reff$, and $2\reff$. Again the qualitative trend is the same as in
the other panels, with minimum value well below unity for $\xi
>1$. Note that for $R=\reff$ and considering the limit of eq. (20) for
$\xi \to \infty$, we obtain for the mass ratio the value $\simeq
1.43$, in perfect agreement with the analogous result for CMZ09
models.

It is interesting to compare  the DM halo
  profile of JJ models in eq. (17) with the NFW profile (Navarro
  et al. 1997), that we rewrite for $r <\rt$ (the so-called
  truncation radius) as
\begin{equation}
\rhoNFW(r)={(\MR -1)\rhon\over f(c)s(\csih +s)^2},\quad f(c)=\ln(1+c) -{c\over 1+c},
\end{equation}
where $\csih \equiv r_{\rm NFW}/\rs$ is the NFW scale-lenght in units of
$\rs$, and $c\equiv\rt/r_{\rm NFW}$: note that in equation above we impose that
the total halo mass $\MD$ is the same as in eq. (17). From the
asymptotic expansion of $\rhoDM$ we already know that $\rhoDM$ and
$\rhoNFW$ at small and large radii cannot in general be similar. Hovever,
in the case of {\it minimum halo} with $\xi\geq 1$, near the center
$\rhoDM$ increases as $1/r$, so that $\rhoDM$ and $\rhoNFW$ 
can be made indentical for $r\to 0$ with the additional choice
\begin{equation}
\csih =\sqrt{{\xi\over 2f(c)}}.
\end{equation}
Therefore once a specific JJ minimum halo model is considered and a
radial range fixed, eqs. (21)-(22) allow to determine the best-fit NFW
profile with same total mass and central density profile of $\rhoDM$
by tuning the value of $c$. For example, after a simple
``trial-and-error'' exploration, we found that over a range extending
out to $\simeq 4-8\reff$, a ``best-fit'' NFW profile can be made to
agree with a minimum halo $\rhoDM$ with $\xi$ in the range $\simeq
2-5$, with deviations $< 10-20\%$ (at large radii), and $< 5\%$ inside
$\simeq 4R_{\rm e}$, adopting $c$ in the range $\approx 10-20$, and
resulting $r_{\rm NFW}$ in the range $\approx 0.9 - 1.5\reff$.

\section{The phase-space  distribution function}

Having established the {\it structural} limitations of the models,
before solving the Jeans equations, it is useful to discuss some basic
property of the phase-space distribution function (hereafter DF) of JJ
models, in order to exclude {\it dynamically} inconsistent
combinations of parameters (i.e., choices that would correspond to a
somewhere negative DF). Fortunately, as discussed extensively in CP92
(see also Ciotti 1996, 1999), it is possible to obtain lower bounds
for the OM anisotropy radius as a function of the density slope and
the total mass profile, without actually recovering the DF, which is
in general impossible in terms of elementary functions.  More
specifically, in CP92 a simple theorem was proved regarding the
necessary and sufficient limitations on $\ra$ in multi-component OM
models. We also recall that the CP92 result has been shown to be just
a very special case of a class of important and more general
inequalities connecting the local density slope and the anisotropy
profile in consistent spherical models (the so-called Global Density
Slope - Anisotropy Inequality, GDSAI, e.g., see de Bruijne et
al. 1996, An \& Evans 2006, Ciotti \& Morganti 2009, 2010ab, van Hese
et al. 2011).

Thus, following the standard nomenclature (e.g., Binney \& Tremaine 2008), we assume
for the stellar component a DF with the OM parameterization
\begin{equation}
f=f(Q),\quad Q\equiv \Er - {J^2\over 2\ra^2},
\end{equation}
where $\Er =\PsiT-v^2/2$ and $J$ are the relative energy and
angular momentum modulus of each star (per unit mass), respectively,
and $\Psi =-\Phi$ is the relative potential; moreover the DF is
truncated as $f(Q)=0$ for
$Q<0$.  As a central BH of mass $\Mbh$ is added at the center of the
galaxy, the total (relative) gravitational potential is
$\PsiT=\Psig+G\Mbh/r$, and from eq. (13)

\begin{equation}
{\PsiT (r)\over\Psin}\equiv
\psi(s) = 
{\mu\over s}+{\MR\over\xi}\ln {\xi+s\over s},
\quad \mu={\Mbh\over\Ms}.
\end{equation}
As well known the radial ($\srad$) and tangential ($\stan$) components of the
velocity dispersion tensor in OM models are related as
\begin{equation}
\beta(r)\equiv{1-{\stan^2(r)\over{2\srad^2(r)}}}={r^2\over{r^2+\ra^2}},
\end{equation}
so that the fully isotropic case is obtained for $\ra\to\infty$, while
for $\ra=0$ the galaxy is supported by pure radial orbits.  For
finite values of $\ra$, the velocity dispersion tensor becomes
isotropic for $r\to 0$ (in practice for $r<\ra$), and fully
radially anisotropic for $r\to\infty$ (in practice for $r>\ra$).
Introducing the augmented density
\begin{equation}
\varrho(r)\equiv \rhos(r)\left(1+{r^2\over\ra^2}\right),
\end{equation}
the phase-space DF of the stellar component can be recovered from the
inversion integral
\begin{eqnarray}
f(Q)&=&{1\over\sqrt{8}\pi^2}{d\over dQ}
     \int_0^{Q}{d\varrho\over d\PsiT}
     {d\PsiT\over\sqrt{Q-\PsiT}}\cr 
     &=&{1\over\sqrt{8}\pi^2}
     \int_0^{Q}{d^2\varrho\over d\PsiT^2}
     {d\PsiT\over\sqrt{Q-\PsiT}};
\end{eqnarray}
an analogous expression holds for the DF of the isotropic DM halo,
obtained by using $\varrho=\rhoDM$, and 
$\ra =\infty$ in eq. (23). 

In the integral above it is intended that $\varrho$ is expressed in 
terms of $\PsiT$, and the second identity follows from integration by 
parts when considering spatially untruncated profiles such those of JJ 
models. Note that the OM inversion for the CMZ09 model is somewhat
different (see eqs. [19]-[28] therein, and relative discussion),
because for these latter models $Q$ is not defined in terms of the
relative potential (the potential in eq. [16] is purely logaritmic and
so diverges both $r\to 0$ and $r\to \infty$, making the introduction
of the relative potential useless), and $f(Q)$ is {\it not} truncated as a
function of $Q$.

In Sects. 3.1 and 3.2, after a general discussion about
the limitations on the $\ra$ imposed by the request of phase-space
consistency, i.e., $f(Q)\geq 0$ over the accessible phase-space, we
will see how far we can proceed analytically in the recovery of the DF
of JJ models with central BH.

\subsection{Necessary and sufficient conditions for consistency}

Following CP92 a {\it necessary condition} for the positivity of the
DF of each of the mass components of JJ models (stars or DM) in the
total (galaxy plus central BH) potential is that
\begin{equation}
{d\varrho (r)\over dr}\leq 0\quad{\rm[NC]}:
\end{equation}
this condition is independent of the behavior of
the other density components of the system. A {\it weak sufficient
  condition} for consistency is obtained by requiring that the 
derivative inside the last integral in eq. (27) be positive. Also this
condition can be expressed as a function of radius as
\begin{equation}
{d\over dr}\left[{d\varrho(r)\over dr}{r^2\over\MT (r)}\right]\geq 0,
\quad{\rm [WSC]},
\end{equation}
where the total mass profile is given by
\begin{equation}
\MT(r)=\Mg(r)+\Mbh,
\end{equation}
and $\Mg (r)$ is given in eq. (13).  Therefore, a model failing
eq. (28) is certainly inconsistent, while a model obeying eq. (29) is
certainly consistent.  It follows that the true boundary in the
parameter space separating consistent and inconsistent models - that
in general can be only determined by direct inspection of the DF - is
``bracketed'' by the NC and WSC limits.
\begin{figure}
\includegraphics[height=0.4\textheight,width=0.5\textwidth]{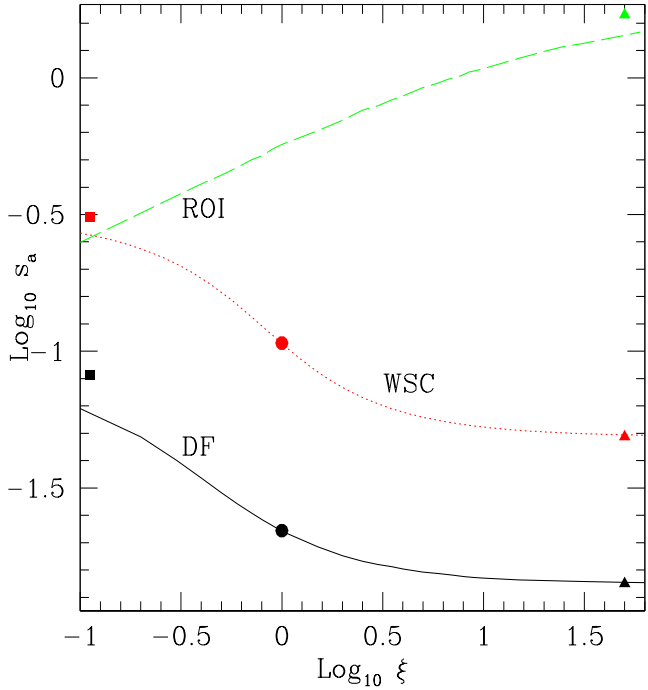}
\caption{ Different limitations on the anisotropy radius $\sa=\ra/\rs$
  of the stellar component of JJ models, as a function of
  $\xi=\rg/\rs$. All the shown results refer to $\mu=0$, i.e. in
  absence of the central BH, when the critical values of $\sa$ are
  rigorously independent of $\MR$. The black solid line and the red
  dotted lines represent the minimum value of $\sa$ obtained directly
  from the DF and from the WSC, respectively, while the green dashed
  curve represents the fiducial lower limit of $\sa$ to prevent the
  onset of Radial Orbit Instability. The triangles show the
  corresponding values for the CMZ09 model ($\sa\simeq 0.0141$,
  $\sa\simeq 0.0487$, and $\sa\simeq 1.78$), and the squares for the
  BH dominated case ($\sa\simeq 0.082$, see Appendix C, and $\sa\simeq
  0.31$, see Appedix A). Finally, the circles correspond to the single
  component (i.e., $\xi =1$) Jaffe model ($\sa\simeq 0.02205$, see
  Appendix C, and $\sa\simeq 0.1068$, see Appendix A).}
\label{fig:wsc}
\end{figure}

Before embarking in the analysis of JJ models, some preliminary
consideration is in order.  First, about the effect of the central BH
on consistency. From eqs. (29) and (30) it follows quite easily that
{\it if} 1) the component is consistent for $\Mbh =0$, and 2) $d (r^2
d\varrho/dr)/dr\geq 0$, {\it then} the model with central BH is
certainly consistent. Note that point 2) is nothing else than the WSC
for the considered density profile interpreted as a tracer in the
gravitational field of the central BH itself; we will use this result
in the following discussion. A second consideration is about the
effect of anisotropy. When dealing with OM anisotropic systems, the
investigation of the NC and WSC, and the study of the DF positivity,
lead to cosider inequalities of the kind
\begin{equation}
F+{G\over\sa^2}\geq 0, \quad \sa\equiv{\ra\over\rs},
\end{equation}
that must hold over the domain $\cond$ spanned by the functions'
argument.  In practice, the functions $F$ and $G$ are functions of $r$
(in the case of the NC and WSC) or functions of $Q$ (in the case of
the DF). From inequality (31) it follows that all OM models can be divided in two
families.  When $F$ is nowhere negative over $\cond$ (e.g., in the
case of a consistent isotropic DF), consistency in the anisotropic
case is obtained for
\begin{equation}
\sa \geq \sam \equiv \sqrt{{\rm max}\left[0,{\rm sup}_{\cond}\left(-{G\over F}\right)\right]}.
\end{equation}
If $G$ is also positive over $\cond$, then $\sa =0$ and the system can
be supported by radial orbits only.  In the second case $F$ is
positive only over some proper subset $\condp$ of $\cond$, and
negative (or zero) over the complementary subset $\condm$.  If $G <0$
somewhere\footnote{ In Ciotti (1999) and Ciotti (2000) it is
  erroneously stated that the model is inconsistent if $G<0$
  {\it everywhere} on $\condm$.  All the results presented therein are
  however correct.}  on $\condm$, then the condition (31) cannot be
satisfied and the model is inconsistent.  If $G \ge 0$ on $\condm$ one
must consider not only the lower limit $\sam$ in eq. (32) evaluated
over $\condp$, but also the condition
\begin{equation}
\sa \le \sap=\sqrt{{\rm inf}_{\condm}\left(-{G\over F}\right)},
\end{equation}
and consistency is possible only if $\sam<\sap$.  Summarizing, if
$F\geq 0$ then $\sa\geq\sam$ for consistency. If $F\leq 0$ over
$\condm$ and $G\geq 0$ there, then the inequality
$\sam\leq\sa\leq\sap$ must be verified. Finally, if over $\condm$ the
function $G <0$ somewhere, or $\sap <\sam$, then inequality (31)
cannot be satisfied and, in case of a DF analysis, the model must be
rejected as inconsistent.

The first application of eqs. (28)-(29) to JJ models concerns the
consistency of the DM halo.  Following the similar analysis in CMZ09,
for simplicity we restrict to the isotropic case, and then eq. (28)
shows the equivalence of the request of monotonicity of $\rhoDM$
(Sect. 2.2) with the NC for a consistent DM halo.  Of course, the
restriction to isotropic case is quite arbitrary, as the virialized
end-states of $N$-body collapses are invariably characterized by some
amount of radial anisotropy (e.g., van Albada 1982; Nipoti, Londrillo
\& Ciotti 2006), but for the present illustrative purposes this
assumption is fully justified.  The WSC for a fully isotropic DM halo
is worked out analytically in Appendix A.  In particular, when
restricting to the case of no central BH ($\mu = 0$) we found, quite
surprisingly, that the condition imposed by the WSC to the halo {\it
  is nothing else than the limit (18) imposed by positivity and
  monotonicity}.  It remains to discuss the effect of a central BH.
Following the argument after eq. (30), it is not difficult to show
(Appendix A) that the addition of the central BH in case of isotropy
{\it reinforces} consistency, i.e., a DM halo that is consistent in
absence of central BH, it is certainly consistent when a BH is added.
Taken together, the two results above and those in Sect. 2.2 shows
that the isotropic DM halo of JJ models with central BH, once {\it
  positivity only} of $\rhoDM$ is assured, automatically satisfies the
NC and WSC conditions, and so it is supported by a nowhere negative
phase-space DF.

We now move to the more interesting case of the NC and WSC for the
stellar component of OM anisotropic JJ models. First, we recall that
NC of Jaffe models just reduces to have $\sa\geq 0$ while, from the
solution of a cubic equation the WSC for the one-component Jaffe model
gives $\sa\geq\sam\simeq 0.1068$ (Ciotti 1999), marked by the red
solid red circle in Fig. 2.  Second, in Appendix A we show that the
WSC of the stellar component of JJ models is always in the case
described by eq. (32), i.e. only $\sam$ exists. However, the function
at the r.h.s. of eq. (32) in the general case is sufficiently
complicated that only a numerical study is feasible. In any case, as
in the next Section we will determine the {\it exact} limit on $\sa$
obtained from the DF, here we just restrict to the case $\mu=0$. The
resulting eq. (A8) is much simpler than the general one, and in
particular $\sam=\sam(\xi)$, i.e. when $\mu =0$ the limit on
anisotropy is independent of $\MR$ (dotted red line in Fig. 2).  The
red triangle at $\sam\simeq 0.0487$ marks the position of the WSC
limit for the CMZ09 model obtained by solving a cubic equation, and
that as expected is in accordance with the value of the red line for
$\xi\to\infty$. At the opposite limit we have the BH dominated case
(see Appendix A), with $\sa\simeq 0.31$ marked by the red square,
coincident with the value of the red line for $\xi\to 0$, when the
total potential becomes that of a central point mass.  In practice,
from the arguments after eq. (30), we have now proved that the stellar
component of JJ models with central BH and OM anisotropy is certainly
consistent for $\sa > 0.31$, independently of the mass of the central
BH and of the DM halo total mass and scale-lenght.  We conclude this
introductory analysis by noticing the fact that for JJ models, the
presence of a diffuse halo appears to increase the model ability to
sustain radial anisotropy, while for concentrated halos the
consistency of the stellar distribution requires a more isotropic
velocity dispersion tensor, as already found in other two-component OM
models (Ciotti 1996, 1999, CMZ09).

\subsection{Explicit phase-space DF}

With the introduction of the dimensionless potential
$\psi=\PsiT/\Psin$ and augmented density $\tilde\varrho=\varrho/\rhon$
from eqs. (24) and (26), respectively, eq. (27) writes
\begin{eqnarray}
f(q)&=&{\rhon\over\sqrt{8}\pi^2\Psin^{3/2}}
     \int_0^{q}{d^2\tilde\varrho\over d\psi^2}{d \psi \over\sqrt{\psi -q}}=\cr
    &&{\rhon\over\sqrt{8}\pi^2\Psin^{3/2}}
       \left[U(q)+{V(q)\over\sa^2}\right],
\end{eqnarray}
where $q\equiv Q/\Psin$. From eqs. (23)-(24) it follows that $0\leq
q\leq\infty$.

In Appendix B we show that it is possible to invert eq. (24) and
express analytically the radius as a function of the relative total
potential $\PsiT$ by using the Lambert-Euler $W$ function, obtaining
\begin{equation}
s(\psi)= {\xi\over {\MR W\over\mu}-1},\quad 
W=W\left(0, {\mu e^{{\xi\psi +\mu\over\MR}}\over\MR} \right),
\end{equation}
where $W(0,z)$ is one of the two branches of the real determination of
the complex function $W$. In absence of the central BH we have
$\PsiT=\Psig$ and it can be shown that eq. (35) reduces to the elementary
function
\begin{equation}
s(\psi)={\xi\over e^{\psi\xi/\MR}-1},
\end{equation}
in agreement with the solution of eq. (24) with $\mu =0$.  With the
substitution (35) in eq. (26) we finally obtain the expression for
$\tilde\varrho (\psi)$ to be used in eq. (27). The derivatives inside
the integral are evaluated from the exact relation in eq.  (B3). We
note that the field of application of the $W$ function to physical
problems is rapidly expanding (e.g. see Valluri et al. 2000, Cranmer
et al. 2004, Ciotti \& Bertin 2005 for an application to
self-consistent toroidal structures, Veberic 2012, Waters \& Proga
2012, Herbst 2015, Ciotti \& Pellegrini 2017 for the solution of
isothermal accretion on BHs at the center of galaxies).

In CMZ09 it is shown that for the stellar Jaffe model embedded in a
total singular isothermal density profile, and in absence of the
central BH, the functions $U$ and $V$ can be expressed as simple
linear combinations of exponentials and polylogarithms.  Here, not
surprisingly, the functions $U$ and $V$ cannot be expressed in terms
of known functions, even in absence of the central BH. However, it is
interesting to notice that in case of a dominant central BH (in
practice, sufficiently near to the center), the function $f(q)$ can be
expressed by using simple functions (Appendix C).

We now determine numerically the lower limit for consistency of $\sa$
by inspection of the functions $U$ and $V$.  Note that in absence of
the central BH ($\mu=0$), the variable $q$ in eq. (34) can be further
scaled as $\qt=q/\MR$, while a factor of $\MR^{-3/2}$ appears in the
functions $U$ and $V$, as shown in eq. (C4). In particular, for these
models without BH the position of the maximum in eq. (32) depends on
$\qt$ (and so in terms of $q$ scales linearly with $\MR$), but the
value of $\sam$ is independent of $\MR$. The same situation occurs in
the CMZ09 models, and in the extreme case of a BH dominated JJ model,
where $\qt=q/\mu$ (Appendix C, eqs. [C4]-[C5]), and the scaling
arguments above apply to the DF with $\MR$ replaced by $\mu$.  It is
numerically found that $U\geq 0$, so that eq. (32) applies and only
$\sam$ exists: the solid line in Fig. 2 shows $\sam(\xi)$ determined
by the DF in absence of the central BH, for comparison with the other
curves presented.  Notice how the shape of the critical consistency
curve parallels the WSC condition (red dotted line), and how there are
consistent models failing the WSC. The black circle at $\xi =1$ marks
the value of the minimum value $\sa\simeq 0.02205$ for the OM
one-component Jaffe model (Appendix C).  From the figure it i apparent
how the effect of a concentrated DM halo reduces the ability of the
stellar component to sustain radial orbits, while the opposite happens
for models with $\xi >1$. As an independent test of the derived DF,
the black triangle indicates the limit value of $\sa\simeq 0.0141$
obtained in CMZ09 by numerical inspection of the DF (coincident, as
expected, with the limit value of the curve for $\xi\to\infty$), and
the black square the value of the BH dominated case $\sa\simeq 0.082$,
coincident with the limit of the curve for $\xi\to 0$.

\begin{figure}
\includegraphics[height=0.35\textheight,width=0.45\textwidth]{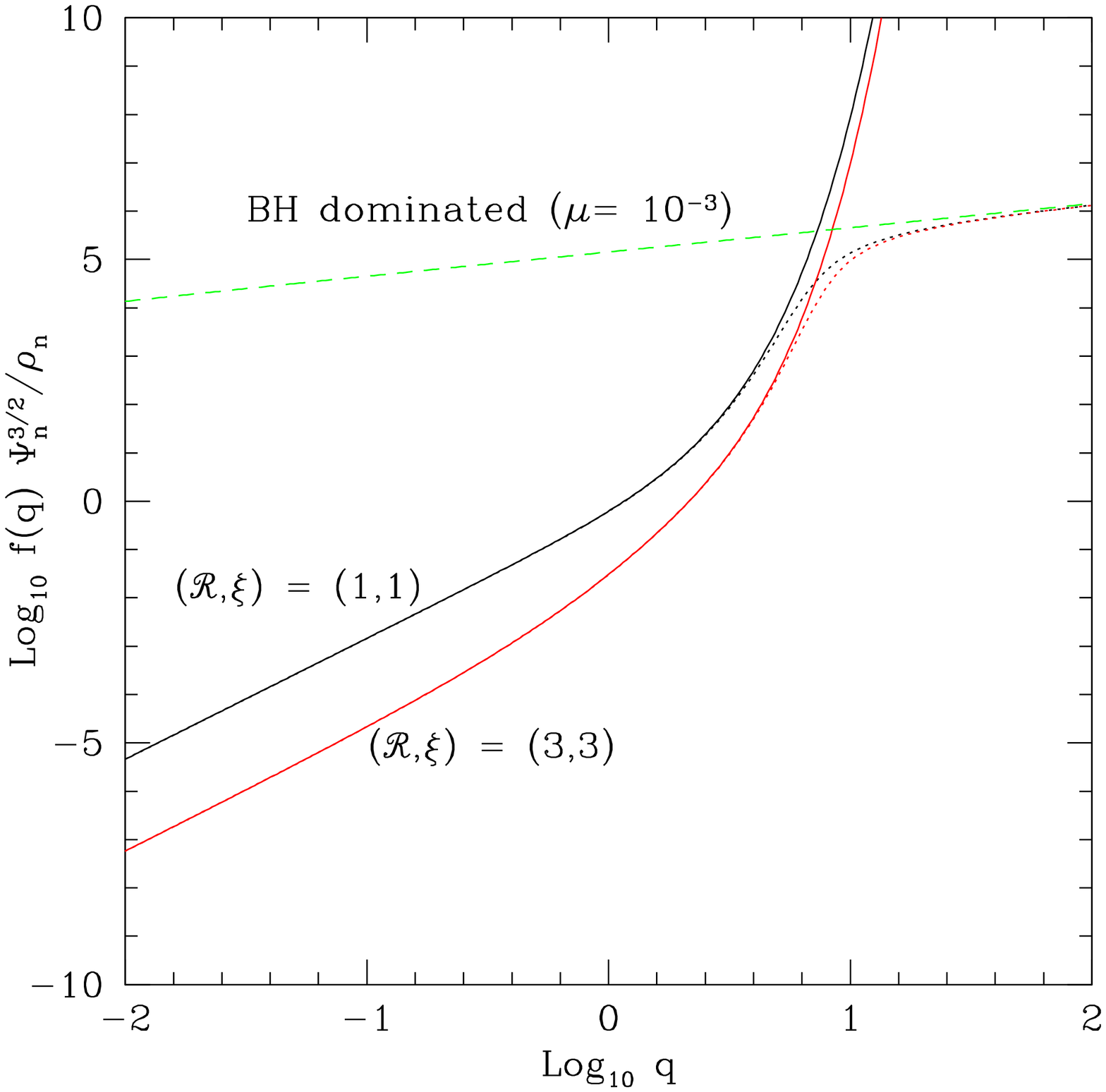}\\
\includegraphics[height=0.35\textheight,width=0.45\textwidth]{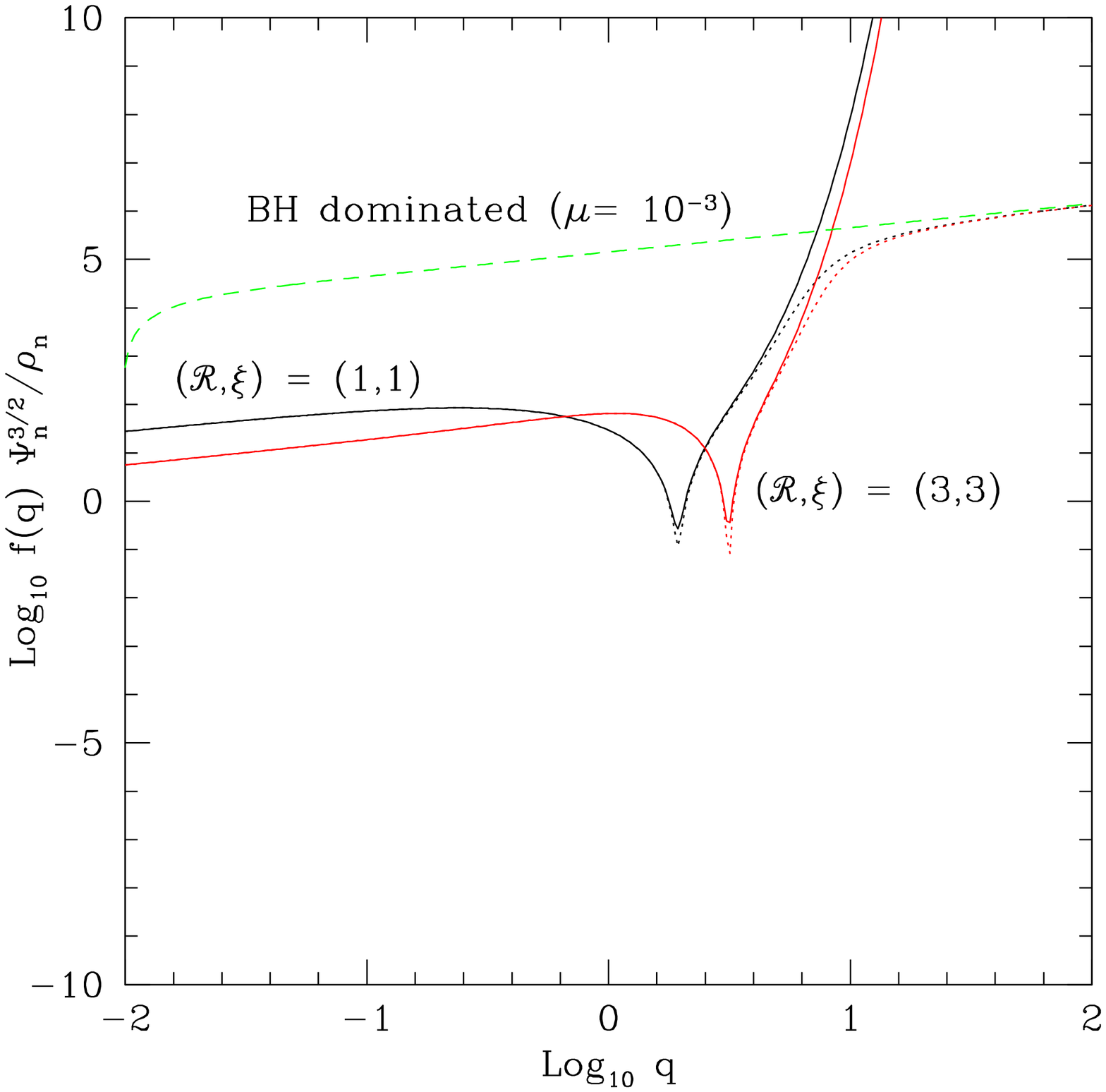}
\caption{The phase-space DF (normalized to $\rhon/\vc^3\sqrt{8}\pi^2$) 
 of the stellar component of $\gamma=1$
 (top) and $\gamma=2$ (bottom) models embedded in a dark matter halo 
 so that the total density profile is proportional to $r^{-2}$. Solid 
 lines refer to the case of a fully isotropic stellar component,
 dotted lines to intermediate values of the (normalized) anisotropy 
 radius $\sa$ ($1$ for $\gamma=1$ and $0.1$ for $\gamma=2$), and 
 finally the dashed lines to a value of $\sa$ very near to the 
 critical value for consistency.}
\label{fig:DF}
\end{figure}

In Fig. 3 the DF of the stellar component of a selection of
representative JJ models is presented, in the isotropic (top panel)
and anisotropic (bottom panel, $\sa=0.1$) cases. In both cases the DFs
are shown with and without the effect of the central BH (with $\mu =
10^{-3}$), and for illustration, also the BH dominated DF (green
dashed line) is shown. It is clear how at high (relative) energies the
DF of JJ models with central BH is perfectly described by the BH
dominated DF. Also, it is apparent how at high energies the isotropic
and anisotropic DFs for models with the same structure are almost
identical, a property of OM anisotropy parameterization leading to
almost isotropic models in the central regions. It is also important
to note how the DFs of models {\it without} the central BH are higher
at high energies than in the analogous models {\it with} the central
BH. Also, notice how models with heavier and more extended halos and
so with higher velocity dispersions at large radii (Fig. 4, top panel) at low
relative energies have a lower DF. The physical reason of this
behavior is due to the fact that, qualitatively, the phase-space DF is
inversely proportional to the cube of velocity dispersion (because the
integral over the velocity space, at fixed position, must reproduce
the same value of the local density), so that, empirically, high
velocity dispersions corresponds to low values of the DF. This is
particularly apparent in the BH dominated case, with a low DF at high
energies and a high DF at low energies. This is also confirmed by the
low-energy tail of the DF, which is higher in the anisotropic
cases. In fact, from eqs. (25) and (35) it follows that for
$r\to\infty$, the total velocity dispersion profile is proportional to
$({\cal A}+\sa^2{\cal I})/r^2$, i.e., it is lower for smaller values
of $\sa$. Finally, notice how orbital anisotropy produces a drop of
the DF at intermediate energies, with a depression that would be of
increasing depth for decreasing values of $\sa$, finally leading to an
inconsistent DF. The curves relative to the anisotropic cases are very
similar to the analogous curves in Ciotti \& Lanzoni (1997, Fig. 2),
and C99 (Figs. 2 and 3) and CMZ09 (Fig. 3), revealing the common
qualitative behavior of OM anisotropic DFs near the consistency limit,
i.e. the fact that the inconsistency manifests itself in general at
intermediate energies (see also Ciotti \& Morganti 2008 for a
discussion).

\section{Jeans equations with OM anisotropy}

The Jeans equations for spherical systems with general (radial or
tangential) anisotropy has been discussed in Binney \& Mamon (1982),
and in the OM case the solution can be written as
\begin{eqnarray}
\rhos (r)\srad^2(r)&=&{G\over r^2 +\ra^2}
             \int_r^{\infty}\rhos(r) \MT(r) \left(1+{\ra^2\over r^2}\right)dr\cr
            &=&\rhon\Psin {{\cal{A}}(s)+ \sa^2 {\cal{I}}(s)\over s^2+\sa^2},
\end{eqnarray}
where $\MT(r)$ is given in eq. (30),  and the two radial functions
\begin{equation}
 {\cal{I}} =\MR\,\It (s) +\mu\,\Ibh (s), \quad {\cal{A}} =\MR\,\At
 (s)+\mu\,\Abh (s),
\end{equation}
are the isotropic and purely radial anisotropic components of the
velocity dispersion tensor, respectively.  In the formula above the
dimensionless mass factors, $\MR$ and $\mu$, have been explicitely
factorized. For $\sa\to\infty$ we obtain the solution of the fully
isotropic case, while for $\sa=0$ we obtain the purely radial case.

\subsection{The velocity dispersion profile}

The integration of eq. (37) is elementary. In fact it is formally
equivalent to an integration already performed, for the different
class of two component Jaffe models in CLR96. where the OM Jeans
equation is integrated for a stellar Jaffe distribution, superimposed
to Jaffe DM halo of total mass $\MD$, and length scale
$r_{\rm{DM}}$. Therefore in CLR96 the combined contribution of the
stars and of the DM potential to the stellar velocity dispersion
profile is given by the sum of two different expressions. Here,
instead, only {\it one} integration is required because the {\it
  total} potential is assigned and, in practice, with a suitable
renaming of parameters, the formula in CLR96 for the DM halo
contribution could be used.  However, as we now consider also the
effect of the central BH, not included in the models in CLR96 and
CMZ09, we give the full set of formulae in homogeneous notation.

For the isotropic component
\begin{equation}
\It=\cases{
         \displaystyle{{\ln(\xi+s)\over\xi^3(\xi-1)^2}+{(3\xi-4)\ln(1+s)\over
             (\xi-1)^2}-{(3\xi^2+2\xi+1)\ln s\over\xi^3} }
         \cr\cr
         \displaystyle{-{2s^2(3\xi^2-\xi-1)+s(3\xi+2)(\xi-1)-\xi(\xi-1)\over 2\xi^2(\xi-1)s^2(1+s)}},
         \cr\cr\cr
         \displaystyle{-{(6s^2+6s-1)(2s+1)\over 2s^2(1+s)}-6 \ln
           {s\over 1+s}},
         }
\end{equation}
where the first expression holds for $\xi \ne 1$, and the second for
$\xi=1$. As expected the two expressions agree with eqs. (A11) and
(A5) in CLR96, respectively\footnote{Due to a typo, the sign of the
  terms inside the square brackets of eq. (A11) in CLR96 should be,
  from left to right, plus, plus, minus, minus.}.  The contribution of
the BH to the stellar isotropic velocity dispersion profile is given
by
\begin{equation}
\Ibh = {12s^3+6s^2-2s+1\over 3s^3(1+s)}+4 \ln {s\over 1+s}.
\end{equation}
Note that this expression could be formally obtained also by
considering the limit for $\xi\to 0$ of the function $\It$, because
from eq. (2) $\PhiJ$ for fixed $r$ and $\rj\to 0$, becomes the
potential of a point mass.

For the anisotropic part we have
\begin{equation}
\At=\cases{
            \displaystyle{{\ln(\xi+s)\over \xi
                (\xi-1)^2}+{(\xi-2)\ln(1+s)\over (\xi-1)^2}-{\ln
                s\over \xi}-{1\over (1+s)(\xi-1)}},
         \cr\cr\cr
         \displaystyle{-{2s+3\over 2(1+s)^2}-\ln {s\over 1+s}},
         }
\end{equation}
where the first expression holds for $\xi \ne 1$ and the second for
$\xi =1$, and they agree with eqs. (A10) and (A4) in CLR96,
respectively. The contribution of the central BH to the anisotropic
stellar velocity dispersion profile is
\begin{equation}
\Abh ={1+2s\over s(1+s)}+2\ln {s\over 1+s},
\end{equation}
and again it is simple to prove that $At=\Abh$ for $\xi\to 0$.
Following eq. (15), we also verified eqs. (39) and (41) considering
the limit for $\xi \to \infty$ of the functions $ \xi \It$ and $ \xi
\At$, and recovering eqs. (C2)-(C3) in CMZ09 evaluated for $\gamma=2$.

An insight of the behavior of $\srad$ can be obtained by considering
the expansion for $r\to \infty$ and $r\to 0$ of the obtained formulae.
We begin with the outer galaxy regions.  A simple expansion of the
functions ${\cal{A}}$ and ${\cal{I}}$ shows that for $r\to\infty$ (in
practice, for $r >>\rs$) the leading order term is the same for the
galaxy as for the BH, with
\begin{equation}
\It\sim\Ibh\sim {1\over 5s^5}+O(s^{-6}), 
\end{equation}
\begin{equation}
\At\sim\Abh \sim {1\over 3s^3}+O(s^{-4}).
\end{equation}
The coincidence of the leading term is just due to the fact that for
$r\to \infty$ the cumulative mass profile in eq. (13) converges to the
total galaxy mass, and for the Newton's theorem this leads to the same
contribution to the velocity dispersion as that of a central mass
$\Mg$.  Following the same approach adopted in CMZ09 we combine
eqs. (43)-(44), and the leading term of $\srad$ in eq. (37) for $r\to
\infty$ is obtained, for arbitrary value of $\ra$, by retaining the
leading order term of the expansion of the much simpler expression
\begin{equation}
\rhos(r) \srad^2(r) \sim \rhon\Psin (\MR+\mu) {5s^2+3\sa^2\over 15s^5 (s^2+\sa^2)}.
\end{equation}
In the case of finite $\sa$ we have $\srad^2 \propto 1/(3s) $, while
in the fully isotropic case $\srad^2 \propto 1/(5s)$: as expected, the
isotropic $\srad$ is lower than in case of finite $\ra$, when the
outer regions become populated by radial orbits only. As expected
eq. (45) agrees with the analogous expression obtained for the
two-component model briefly discussed in Sect. 4.4 of CMZ09 (eq. [40]
therein, for $\gamma=2$ and for a dominant DM halo). This is at
variance with the behavior of the genuine CMZ09 model, where for
$r\to\infty$
\begin{equation}
\rhos(r)\srad^2(r)\sim\rhon\vc^2{2s^2+\sa^2\over 4s^4(s^2+\sa^2)}.
\end{equation}
Therefore, although the full velocity profile of the CMZ09 model is
recovered from the the limit procedure in eq. (15) applied to
eqs. (39)-(41), the limit procedure applied to eq. (45) does {\it not}
converge to the asymptotic expansion of the velocity dispersion
profile in CMZ09. This is due to the fact that for $\xi \to \infty$
and $r\to\infty$ the integral (37) is {\it not uniform} in the
variables $\xi$ and $r$, so that the two limits cannot be in general exchanged.

The other important region for observational and theoretical
works, is the galaxy center: here the velocity dispersion profile is
dominated by the BH contribution. In fact, at the leading order
\begin{equation}
\It \sim {1\over 2\xi s^2}+O(s^{-1}),\quad \At \sim -{\ln s\over \xi}+O(1),
\end{equation}
so that for $r\to 0$ the galaxy contribution to the stellar velocity
dispersion profile is given by
\begin{equation}
\rhos(r)\srad^2(r) \sim {\rhon\Psin\MR\over\xi s^2}
\cases{
            \displaystyle{{1\over 2}}, \quad\sa >0 ;
         \cr\cr
         \displaystyle{-\ln s},
                       \quad\quad\sa =0.
         }
\end{equation}
In particular, if $\ra=0$, the central velocity dispersion diverges as
$\srad^2 \propto -\ln s$, while for all values $\ra>0$ the central
velocity dispersion converges to a finite value, coincident with that
of the isotropic case
\begin{equation}
\srad^2 (0)={\Psin\MR\over 2\xi}.
\end{equation}
This is relevant from the modelistic point of view, as it is well
known that self-gravitating isotropic $\gamma$ models present a
depression of their velocity dispersion near the center with
$\srad(0)=0$, except for the $\gamma=0$ and $\gamma=2$ models (e.g.,
see Bertin et al. 2002 for a general discussion of this phenomenon;
see also Binney \& Ossipkov 2001). Notice that the value of the
central velocity dispersion, in the minimum halo model with $\xi\geq
1$ is, according to eq. (18), {\it independent} of $\xi$, and {\it
  coincident} with that of the purely stellar Jaffe model. This shows
the danger of a ``blind'' use of $\srad(0)$ as a robust indicator of
the actual depth and shape of the galaxy potential well.

For the BH we obtain
\begin{equation}
\Ibh \sim {1\over 3s^3}+O(s^{-2}),\quad \Abh \sim {1\over s}+O(\ln s),
\end{equation}
and the formula analogous to eq. (48) is
\begin{equation}
\rhos(r) \srad^2(r) \sim {\rhon\Psin\mu\over s^3}\cases{
            \displaystyle{{1\over 3}},
                                                     \quad\sa >0 ;
         \cr\cr
         \displaystyle{1},
                       \quad\quad\sa =0.
         }
\end{equation}
As expected, $\srad^2$ diverges as $\mu /r$ for $r\to 0$, and with a
factor of 3 of difference between the fully radially anisotropic case,
and all the other cases with $\sa>0$ in agreement with the general
property of $\srad$ in the central regions of $\gamma$ models with a
BH (e.g., see C96, Baes \& Dejonghe 2004, Baes et al. 2005). We
conclude by noticing that eqs. (48), (49) and (51) are also in
accordance with the analogous quantities for the CMZ09 model
(eqs. [C5]-[C6]) and the two-component models in Sect. 4.4 there
(eq. [44]), and with the results in the spherical (isotropic) limit of
one and two-component oblate power-law models with central BH in
Ciotti \& Bertin (2005, eq. [C3]) and in Riciputi et al. (2005,
eq. [A4]), evaluated for $\gamma=2$
\begin{figure}
\includegraphics[height=0.4\textheight,width=0.5\textwidth]{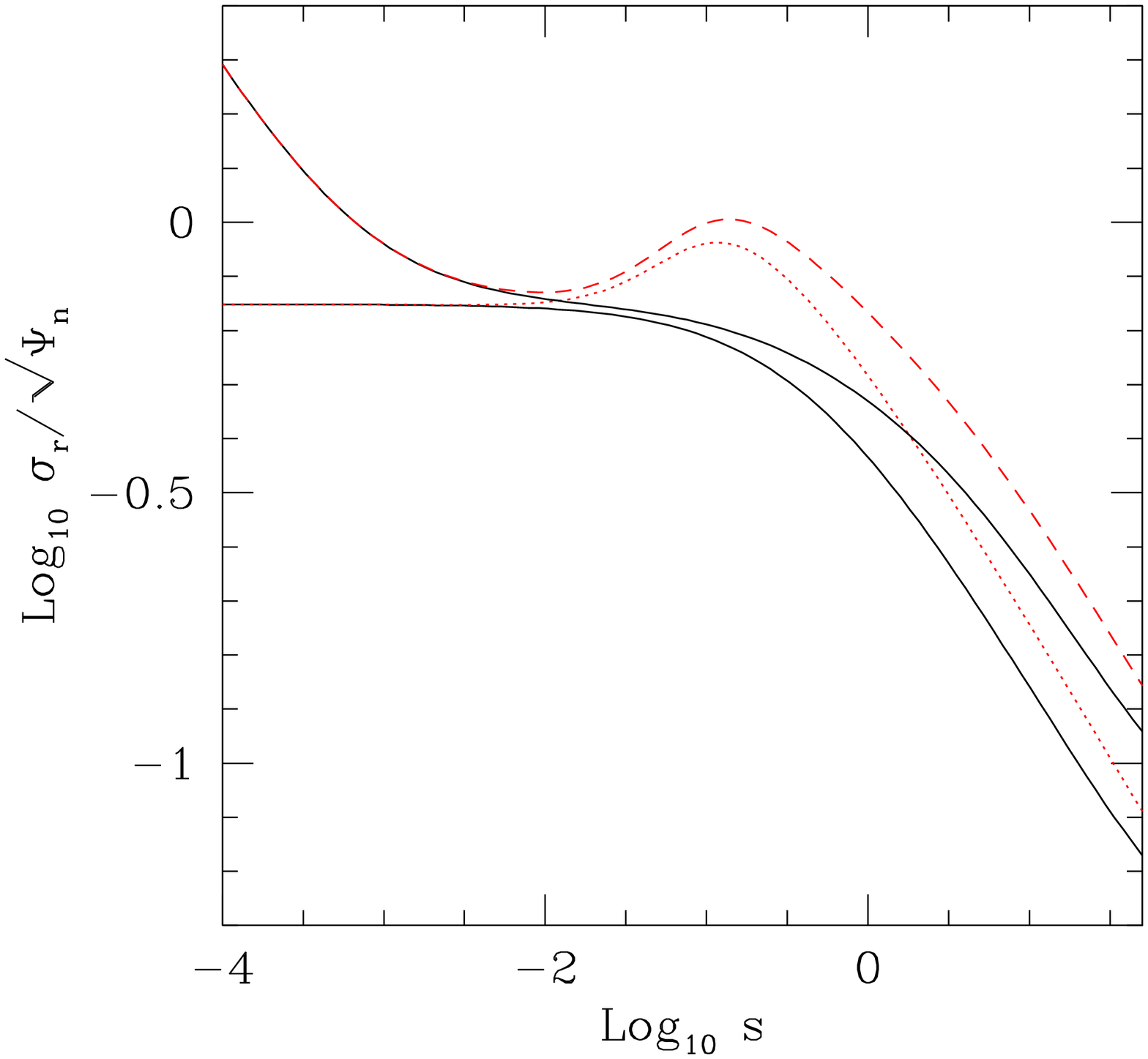}
\includegraphics[height=0.4\textheight,width=0.5\textwidth]{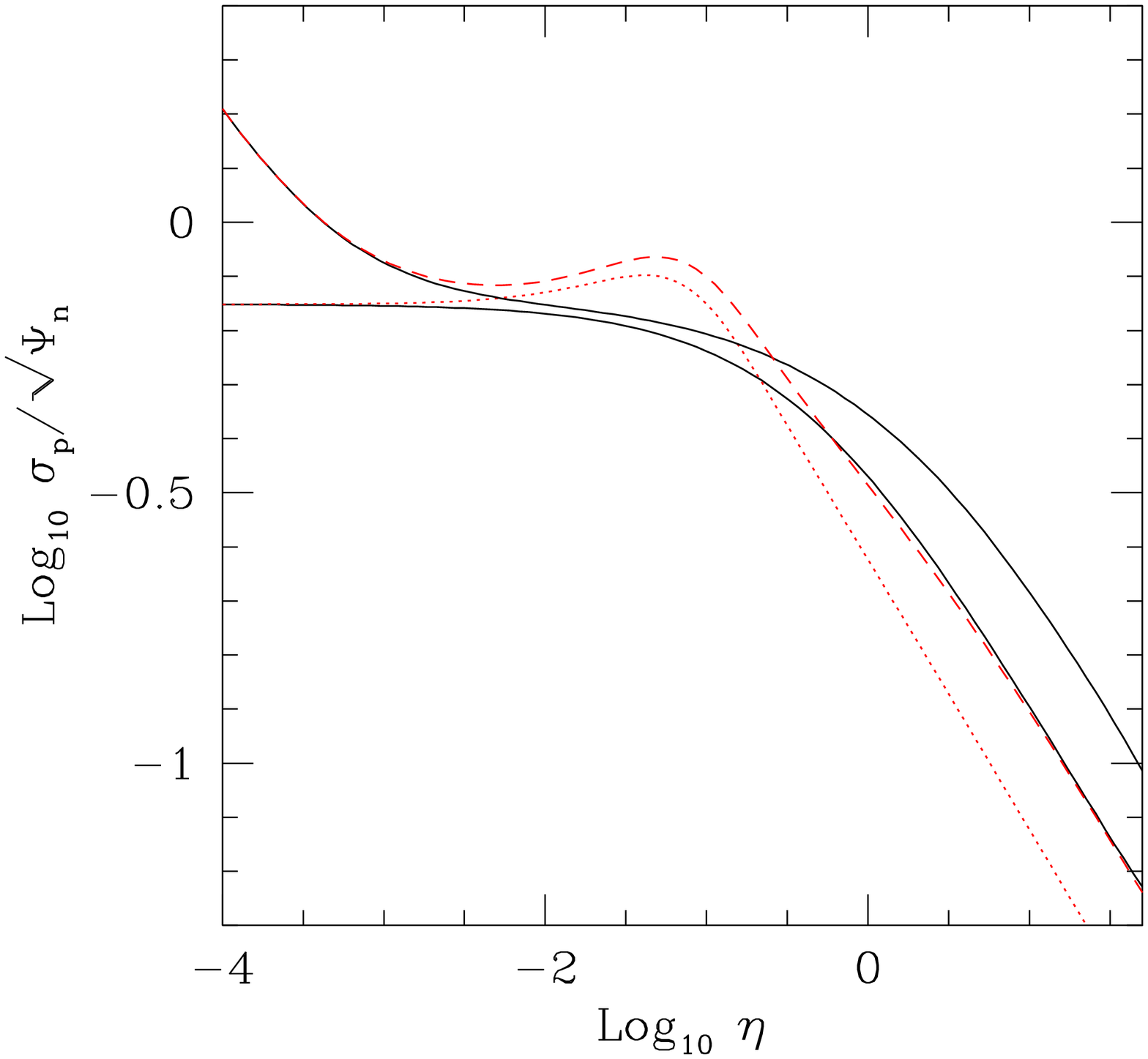}
\caption{Top panel: radial trend of $\srad$ of the stellar component
  of JJ models vs. $s=r/\rs$, in some representative case.  Black
  solid lines refer to the isotropic case for the single component
  Jaffe model ($\MR=1$, $\xi=1$, $\mu=0$), and for a model with
  central BH and a minimum DM halo ($\MR=3$, $\xi=3$, $\mu=10^{-3}$),
  respectively.  Red lines (dotted and dashed) show the profiles for
  the same models but in a quite anisotropic case, with
  $\sa=0.1$. Bottom panel: radial trend of the projected stellar
  velocity dispersion $\sigp$ vs. $\eta=R/\rs$ for the same models in
  the top panel.}
\label{fig:sigma}
\end{figure}
All the relevant properties of $\srad$ described in this Section are
illustrated in Fig. 4 (top panel) by a selection of representative JJ models. In
particular, the effects of the central BH, of the DM halo, and of
orbital anisotropy, can be clealry seen near the center and at large
radii.

\subsection{Projected velocity dispersion}

The projected velocity dispersion profile associated with a general
anisotropy function $\beta(r)$ is given by
\begin{equation}
\Sigma_*(R)\sigp^2(R)= 2\int_R^{\infty}{\left [{1-\beta(r)}
{R^2\over{r^2}}\right ]}{{\rhos(r)\srad^2(r)\,r}\over{\sqrt{r^2-R^2}}}dr,
\end{equation}
(e.g., Binney \& Tremaine 2008), and in the OM case $\beta (r)$ is given
in eq. (25).

Unsurprisingly the projection integral cannot be evaluated
analytically for JJ models in terms of elementary functions.  However,
as for the spatial velocity dispersion profile interesting
informations can be obtained outside the core radius and near the
center.  In practice, in the external regions the stellar and total
density profiles can be approximated as a pure power-law of slope
$-4$. In this region the projection integral can be evaluated for
generic values of $\sa$ and in analogy with eq. (45) the asymptotic
trend with radius of the projected profile can be obtained by
retaining the leading order term of the expansion of
\begin{eqnarray}
\sigp^2(R)&\sim&{8(\MR +\mu)\Psin\over 15\pi\eta}
\left[1+{\eta^4\over 2\sa^2(\sa^2+\eta^2)}\right.\nonumber\\
&-&\left.{\eta^4(2\sa^2+\eta^2)
{\rm archsinh}(\sa/\eta)\over 2\sa^3(\sa^2+\eta^2)^{3/2}}\right],
\end{eqnarray}
where $\eta\equiv R/\rs$. The expression in square parentheses
converges to $1$ in the isotropic case, and to $1/3$ for all finite
values of $\sa$.  The analogous formula for the
CMZ09 limit models is
\begin{equation}
\sigp^2(R)\sim\vc^2{(\sa^2+\eta^2)^{5/2}-\eta^3(2\sa^2+\eta^2)\over
                             4\sa^2(\sa^2+\eta^2)^{3/2}},
\end{equation}
and the same considerations made after eq. (46)  hold.

The case of the central regions is more complicated.  In fact, both
the integral (52) and the projected surface density $\Sigma_*$ (see
eq. [7]) are asymptotically dominated by their integrands for $r\to
0$, so that $\sigp^2$ can be properly defined only as the limit for
$R\to 0$ of the ratio of two diverging quantities. For what concerns
the galaxy contribution, a simple calculation shows that for $\ra >0$
\begin{equation}
\sigp(0)=\srad(0),
\end{equation}
where $\srad(0)$ is given by eq. (49), again in agreement with
eq. (33) in CMZ09 for $\gamma=2$. For $\ra =0$ instead the central
projected velocity dispersion diverges.Therefore, for the stellar
component of JJ models and $\ra >0$, {\it the projected central
  velocity dispersion coincides with the central radial component of
  the isotropic velocity dispersion}. In presence of the central BH,
$\srad$ is dominated by the BH contribution, and so it is the
projected velocity dispersion. With some care, from eqs. (51)-(52) it
can be shown that, from eqs. (51) and (52) and {\it independently of
  the value of $\sa\geq 0$},
\begin{equation}
\sigp^2(R)\sim {2\Psin\mu\over 3\pi\eta}.
\end{equation}
All the relevant properties of $\sigp$ expressed by the formulae in
this Section can be noticed in Fig. 4 (bottom panel), where we show
the projected velocity dispersion profiles for the same JJ models in
the top panel.  In particular Fig. 4 shows a well known consequence of
the OM parameterization, i.e., the fact that the isotropic $\srad$
profiles (black lines) in the outer regions are below those in the
corresponding radially anisotropic cases (red lines), while the
opposite holds for the $sigp$ profiles, due to projection effects on
radial orbits in the outer regions, where the l.o.s. direction is
almost perpendicular to the stellar orbits.

We conclude this Section by noticing that CMZ09 (eq. [39]) briefly
commented on the spatial and projected velocity dispersion of a
two-component galaxy model made by the superposition of a stellar
distribution described by a $\gamma$ model, and a DM halo described by
a Jaffe model.  Of course, when $\gamma=2$ this family reduces to JJ
models in CLR96: in turns it is easy to check the perfect
correspondance of eqs. (49) and (55) with eq. (42) in CMZ09 by
assuming there $\MR\to\infty$ and $\beta=\xi$, when the model becomes
formally identical (in the limiting case of a DM halo ``infinitely
massive'') to JJ models (without central BH).  The formulae (55) and
(56) also agree, as expected, with the projection formulae in the
spherical limit of the ellipsoidal models with $\gamma=2$ in Ciotti \&
Bertin (2005, eqs. [C1] and [C7] therein).

\section{Virial, potential, and  kinetic energies}

Among the several global quantities that are associated with a stellar
system, those entering the Virial Theorem (hereafter VT) are certainly
the most interesting for many observational and theoretical studies
(e.g., Ciotti 2000, Binney \& Tremaine 2008). For the stellar
component of JJ models we have
\begin{equation}
2\Ks \equiv -\Ws =-\Wg-\Wbh,
\end{equation}
where $\Ks=2\pi\int_0^{\infty}\rhos (\srad^2+\stan^2)r^2dr$ is the total kinetic energy of the stars,
\begin{equation}
\Wg= -\int \rhos <\mathbf{x} ,\nabla \Phig>  d^3\mathbf{x}
= - 4\pi G\int_0^{\infty} r\rhos (r)\Mg(r)dr,
\end{equation}
is the interaction energy of the stars with the gravitational field of
the galaxy (stars plus DM), and finally
\begin{equation}
\Wbh=-4\pi G \Mbh \int_0^\infty r \rhos (r) dr,
\end{equation}
is the interaction energy of the stars with the central BH.
For a Jaffe galaxy $\Wbh$ diverges, because the stellar
density profile diverges near the origin as $ r^{-2}$; instead, this
quantity converges for $\gamma$ models with $0\le \gamma<2$. Therefore,
the VT implies that also the volume integral of
$\rhos\sigma_{*\rm{BH}}^2$ diverges near the origin for a Jaffe
galaxy, as can be seen by direct integration of eq. (51)

The contribution of the total galaxy potential to 
$\Wg =\Wss +\Wdm$ (where $\Wss$ is due to the self-interaction of the
stellar distribution, and $\Wdm$ to the effect of the DM halo)
is finite, with the remarkably simple result
\begin{equation}
\Wg=-\Psin\Ms\MR\cases{
            \displaystyle{{\xi-1-\ln \xi\over (\xi-1)^2}},
                                                     \quad \xi \ne 1;
         \cr\cr
         \displaystyle{{1\over 2}},
                       \quad\quad\quad\quad\quad\quad\quad \xi=1,
         }
\end{equation}
and taking the limit as in eq. (15), $\Wg=-G\Ms\vc^2$, in accordance
with eq. (33) in CMZ09. More generally it can be shown that $\Wg$ is a
finite quantity for the stellar component of $\gamma\gamma$ models,
provided $0\leq\gamma < 5/2$ (e.g., for two component Hernquist model,
obtained for $\gamma=1$).  It follows that for this class of models it
is possible to define the (3-dimensional) galactic virial velocity
dispersion as $\sigma^2_{\rm V}=-\Wg/\Ms$: moreover, from eqs. (60),
(55) and (49) the value of $\sigma_{\rm V}^2$ is proportional to the
value of the central projected velocity dispersion $\sigp^2(0)$, and
the proportionality constant is a function of $\xi$ {\it only}: for
$\xi=1$, $\sigma_{\rm V}=\sigp(0)$. We also notice the interesting
behavior of $\Wg$ as a function of $\xi$ in the minimum halo
case. While for increasing $\xi\geq 1$ it follows that $\MR=\Rm=\xi$
increases correspondingly to arbitrarily large values, the
dimensionless coefficient in eq. (60) just increases from $1/2$ for
$\xi=1$ to 1 for $\xi\to\infty$, due to the fact that in minimum-halo
case, more massive halos are necessarily more and more extended, with
a compensating effect on the depth of the total potential.

As well known, in multi component systems the virial energy $W$ of a
given component is {\it not} the gravitational energy of the component
itself in the total potential. For this reason we now calculate
explicitly the different contributions to the potential energy $\Us$
of the stellar component of JJ models, and we also show how to obtain
the expression of $\Wss$ and $\Wdm$ in a simple way.  As for the
interaction energy $\Ws$, also for the potential energy $\Us$ holds the
decomposition
\begin{equation}
\Us=\Ug+\Ubh,
\end{equation}
where 
\begin{equation}
\Ug=\Uss+\Udm={1\over 2}\int \rhos \Phis d^3\mathbf{x}+\int\rhos \Phidm d^3\mathbf{x},
\end{equation}
and
\begin{equation}
\Ubh=\int \rhos \Phibh d^3\mathbf{x}=-4\pi G \Mbh \int r \rhos(r) dr=\Wbh.
\end{equation}
Therefore $\Ubh$ diverges as $\Wbh$. From a well known result, the
self-gravitational energy and the virial self energy of each density
component of a multi-component system coincide, and in our case 
from eq. (60) with $\MR=1$ and $\xi=1$,
\begin{equation}
\Uss = \Wss = -{\Psin\Ms\over 2}, 
\end{equation}
so that we can compute $\Wdm =\Wg -\Wss$ without performing additional
integrations.  The evaluation of $\Udm$ is slightly more complicated,
because in principle it would require to substitute
$\Phidm=\Phig-\Phis$ in the second integral in eq. (60), and therefore
compute two integrals. But we adopt a different strategy, and we
compute the integral
\begin{equation}
\Bg \equiv \int \rhos \Phig d^3\mathbf{x}=-\Psin\Ms\MR
\cases{
            \displaystyle{{\ln\xi\over \xi-1}},
                                                     \quad \xi \ne 1;
         \cr\cr
         \displaystyle{1},
                       \quad\quad\quad\quad\quad \xi=1.
         }
\end{equation}
so that from eq. (60)
\begin{equation}
\Udm= \Bg-2\Uss,
\end{equation}
and finally $\Ug$ is obtained by adding $\Uss$.

Note that $\Bg$ {\it is not} the gravitational energy $\Ug$ of the
stars in the galaxy total potential, $\Ug$. Yet, $\Bg$ is not just an
useful mathematical quantity, but it has an important physical
interpretation, and together $\Ks$ plays a fundamental role in the
theory of galactic winds and X-ray emission of early-type galaxies. In
fact, the energy per unit time to be provided to the ISM of early-type
galaxies (for example by supernova explosion, thermalization of
stellar winds, and AGN feedback) required to steadily extract the mass
losses of stars, injected over the galaxy body at the rate $\dot
\rho_{\rm{inj}}=\alpha(t)\rhos$ is given by
$L_{\rm{grav}}=\alpha(t)|\Bg|$ (e.g, see Pellegrini 2011, 2012,
Posacki et al. 2013). A nice feature of JJ models is that $\Bg$ is
finite and given by a remarkably simple expression, at variance with
the situation of CMZ09 models, where this quantity would diverge, or
other two-component models, where $\Bg$ is given by quite cumbersome
formulae. Therefore JJ models provide a very simple framework to
estimate the energetic of galactic gas flows hosted by X-ray emitting
early-type galaxies.

\subsection{Stability}

Another particularly relevant application of the VT is in the field of model
stability, i.e. the determination of the conditions required to
prevent the onset of the so-called Radial Orbit Instability
(hereafter, ROI). In fact, it is well known that stellar systems
supported by a large amount of radial orbits are in general unstable
(e.g., Fridman \& Polyachenko 1984, and references therein).  A stability analysis is
obviously well beyond the task of this work, but we can obtain some
quantitative information by investigating the value, as a function
of the model parameters, of the stability indicator
\begin{equation}
\Xi \equiv {2 \Krad\over\Ktan} = -{4\over 2+\Wg/\Krad},
\end{equation}
where $\Krad$ and $\Ktan =\Ks -\Krad$ are the total kinentic energes
of the stellar component of JJ models, associated with the radial and
tangential components of the velocity dispersion tensor, respectively,
and the last expression is obtained by evaluating $\Ktan$ from the VT.
Of course, we exclude the effect of the central BH, due to the formal
divergence of the kinetic energy $K_{\rm *BH}$ discussed in previous
Section.  From its definition $\Xi\to1$ for $\sa\to\infty$ (globally
isotropic models), while $\Xi\to\infty$ for $\sa\to 0$ (fully radially
anisotropic models).

Numerous investigations of one-component systems have confirmed that
the onset of ROI is in general prevented by the empirical requirement
that $\Xi < 1.7\pm 0.25$; the exact value of the limit is model
dependent (see, e.g., Merritt \& Aguilar 1985; Bertin \& Stiavelli
1989; Saha 1991, 1992; Bertin et al. 1994; Meza \& Zamorano 1997;
Nipoti, Londrillo \& Ciotti 2002).  Here we are considering
two-component systems, however N-body simulations have shown that the
presence of a DM halo does not change very much the situation with
respect to the one-component systems (e.g., see Stiavelli \& Sparke
1991, Nipoti et al. 2002). In our case, we assume as a fiducial maximum value for
stability $1.7$. 

Note that from eq. (60) and volume integration of eq. (37) with $\mu
=0$, eq. (67) shows that $\Xi$ is independent of $\MR$.  Unfortunately
$\Krad$ cannot be expressed by using elementary functions, so that we
explore numerically the fiducial stability condition $\Xi
(\sa,\xi)=1.7$. In Fig. 2 with green dashed curve we plot the
resulting lower bound for stability $\sa(\xi)$.  It is apparent that
the critical value of $\sa$ increases with $\xi$, and this is due to
the fact that a spatially extended DM halo increases the contribution
to the kinetic energy of the velocity dispersion in the outer parts
that, in the OM case, are radially anisotropic. Therefore, in order to
guarantee stability in presence of an extended DM halo, the permitted
amount of radial orbits must correspondingly reduced, and larger
values of $\sa$ are needed. As a limit case the green triangle marks
the position of the stability indicator for the limit models in CMZ09,
with a limit value of $\sa\simeq 1.78$.  The opposite situation occurs
when the DM halo is more concentrated than the stellar component,
because in this case the velocity dispersion is increased
preferentially in the central regions, that in the OM case are in
practice isotropic, and so a larger amount of radial orbits can be
supported. All these trends nicely agree with those found for
different famillies of one and two-component $\gamma$ models (Ciotti
1996, 1999, see also Carollo et al. 1995).  We finally notice how the
stability criterion requires minimum anisotropy radii appreciably
larger than those obtained from the consistency analysis (see
Sect. 3.2), and so it is likely that the maximally radially
anisotropic models with positive DF, would be prone to develop ROI.

\section{Conclusions}

The family of spherical, two-component galaxy models with the stellar
density distribution described by the Jaffe profile, embedded in a DM
halo such that the {\it total} density distribution is also a Jaffe
profile, is presented.  The DM halo is defined as the difference
between the total and the stellar density distributions.  A BH is
added at the center of to the system, and the dynamics of the stellar
component is described by the Osipkov-Merritt anisotropy profile. The
models are fully determined once the total stellar mass ($\Ms$) and
scale length ($\rs$) are assigned, together with the total-to-stellar
mass ratio ($\MR$), the total-to-stellar scale length ratio ($\xi$),
the BH-to-stellar mass ratio ($\mu$), and finally the anisotropy
radius ($\ra$) of the stellar distribution.  These models represent a
generalization of the CMZ09 models, where the total density profile
was fixed at $r^{-2}$ at all radii. In fact JJ models, while retaining
interesting properties such as a realistic stellar density profile and
a total density profile that can be described an arbitrarily large
radial range by a $r^{-2}$ profile, have a finite total mass, and a
central BH.  At the same time, they still allow for an almost complete
analytical treatment, and several quantities of interest in
observational and theoretical works have remarkably simple explicit
expressions.  The main results can be summarized as follows.

\begin{itemize}

\item After providing a summary of the structural quantities of
  observational interest for JJ models, for the more general family of
  two-component $\gamma\gamma$ models, we derive analitically the
  constraints on $\MR$ and $\xi$ needed to assure positivity and
  monotonicity of the DM halo density distribution.  For a given value
  of $\xi$, the model corresponding to the minimum value allowed for
  $\MR$ is called {\it minimum halo model}.  In JJ models (in which
  the positivity and monotonicity limits coincide), $\MR \ge
  \rm{max}\left(\xi, 1/\xi\right)$.  Near the origin the density
  profile of the DM halo diverges as $\rhoDM\propto r^{-2}$, but in
  the minimum halo model with $\xi >1$ the models are centrally
  ``baryon dominated'', with $\rhoDM\propto r^{-1}$.

\item It is shown that the models presented in CMZ09 are limit cases
  JJ models (in absence of the central BH), and we provide the
  framework to derive all the structural and dynamical quantities of
  the CMZ09 models from those of JJ models.

\item The minimum value of anisotropy radius $\ra$, corresponding to a
  dynamically consistent stellar component (i.e., characterized by a
  nowhere negative DF), is first estimated by using the necessary and
  sufficient conditions given in CP92. It is shown that in absence of
  the central BH the minimum value of $\ra$ so determined is a
  function of $\xi$ only. The critical $\ra$ decreases for increasing
  $\xi$, i.e., as already found in other two-component models, a DM
  halo more extended than the stellar distribution increases the
  ability of the stellar component to sustain radial anisotropy. On
  the contrary, more concentrated DM halos (and in particular a
  central BH), require a more isotropic orbital distribution. The
  preliminary consistency analysis is also performed for the DM halo,
  and it is proved that for isotropic DM halos in JJ models with
  $\mu=0$ the conditions of positivity, monotonicity, and phase-space
  consistency coincide; the addition of a central BH reinforces
  consistency.

\item We then moved to study the phase-space DF for the stellar
  component as given by OM inversion.  We found that for JJ models it
  is possible to express analytically the dependence of radius on the
  total potential in terms of the Lambert-Euler $W$ function, allowing
  for a fast and accurate recovery of the DF. In case of no BH
  $(\mu=0)$ the resulting expression reduces to elementary functions,
  and in the limit case of a dominat BH (or a very concentrated DM
  halo corresponding to $\xi\to 0$), the DF itself can be obtained in
  terms of elementary functions. After presenting a few representative
  cases of DFs, corresponding to different choices of $\MR$, $\xi$,
  and $\ra$, we determined numerically the (minimum) critical value of
  $\ra$ as a function of the model parameters, and we found that the
  obtained curve nicely parallels the bound given by the sufficient
  condition in CP92. We showed that in absence of the central BH, and
  in the case of a dominant BH, the critical $\ra$ depends only on
  $\xi$, and it is independent of $\MR$ and $\mu$, respectively. In
  general, we confirmed that DM halos more extended than the stellar
  component increase the amount of radial anisotropy that can be
  supported by a positive DF, while the opposite happens in case of
  concentrated halos (or in presence of a central BH), again in
  accordance with previous findings relative to different
  two-component OM models. Quite unexpectedly, from the inspection of
  the analytical DF, and by independent numerical verification, we
  found that the single component Jaffe model {\it cannot} support
  purely radial orbits in the OM formulation, as detailed in Appendix
  C.

\item The Jeans equations for the stellar component are solved
  explicitely for generic values of the model parameters in terms of
  elementary functions.  The asymptotic expansions of $\srad$ and
  $\sigp$ for $r\to 0$ and $r\to\infty$ are obtained, and in particular
  it is shown that when $\mu=0$ and for all values of $\ra >0$
  (isotropic case included) $\srad^2(0)=\Psin\MR/(2\xi)$. In this
  case, by asymptotic expansion of the projection integral with
  $\ra>0$, it is also shown that independently of the value of the
  anisotropy radius, $\sigp(0)=\srad(0)$. In presence of the BH, in
  the central regions $\srad^2\propto r^{-1}$ with a coefficient which
  is different for $\ra =0$ or $\ra >0$. In projection, due to a
  compensating effect, $\sigp^2(R)\sim 2\Psin\mu\rs/(3\pi R)$ for
  $\ra\geq 0$.

\item Finally, the analytical expressions of relevant quantities
  entering the Virial Theorem, such as the stellar kinetic energy, the
  virial energy interactions, the potential energies, are derived as a
  functions of the model parameters. With the aid of the obtained
  formulae we determined the minimum value of $\ra$ corresponding to a
  value of $\simeq 1.7$ of the Friedmann-Poliachenko-Shuckman
  instability indicator, so that more anisotropic models are prone to
  the onset of Radial Orbit Instability. Again, in line with previous
  results, the minimum $\ra$ for stability increases for increasing
  $\xi$, and in absence of the central BH its value depends only on
  $\xi$, being independent of $\MR$.

\end{itemize}

We conclude by noting that JJ models, albeit highly idealized, suggest
a few interesting remarks of observational and theoretical character.
For example, after having fixed the properties of the models by using
available observational constraints (e.g., see Negri et al. 2014),
one could use JJ models to investigate how the so called {\it sphere
  of influence} of the BH depends on the galaxy properties and how its
definition is affected by orbital anisotropy. Following a preliminary
study (Ziaee Lorzad 2016), it is natural to define the radius of the
sphere of influence as the distance from the galaxy center where the
quantity
\begin{equation}
\Delta \sigma^2 \equiv 
{\sigma_{\rm *g}^2+\sigma_{\rm *BH}^2-\sigma_{\rm *g}^2\over
\sigma_{\rm *g}^2}=
{\mu\Ibh (r)\over\MR\It (r)},
\end{equation}
reaches some prescribed value (for example 20\%, 50\%, 100\%) as a
function structural and dynamical properties of the galaxy itself.  JJ
models could also be used to obtain some preliminary estimate of
structural/dynamical properties of high-redshift galaxies (e.g., see
Sect. 4.4.1 in Vanzella et al. 2017), thanks to the very simple
expressions of their virial quantities.

Another interesting application of JJ models is in the field of BH
accretion because, as shown in Ciotti \& Pellegrini (2017), it is
possible to solve analytically the generalized isothermal Bondi
accretion problem in Jaffe (or Hernquist) potentials with a central
BH. As the total density profile of JJ models is a Jaffe law, it
follows that for these models we can solve {\it both} the accretion
problem for the gas and the Jeans equations for the stellar component.
Moreover, JJ models allow for the computation of the stellar kinetic
energy, a quantity strictly related to the average temperature of the
ISM in early-type galaxies.  As the gas temperature determines the
location of the Bondi radius, JJ models represent a fully analytical
family of self consistent stellar dynamical-hydrodynamical models,
that will allow to compare the relative position of the sonic radius
and the radius of the sphere of influence as a function of the galaxy
properties.

\section*{Acknowledgments}
We thank the anonymous Referee for useful comments, and John Magorrian
and Silvia Pellegrini for interesting discussions on the models.
L.C. thanks G. Bertin, J. Binney, T. de Zeeuw, W. Evans,
D. Lynden-Bell, D. Merritt and S. Tremaine for useful comments on
Appendix C.  A.Z. acknowledges the Department of Physics and Astronomy
of Padua University, where a preliminary study of JJ models has been
the subject of her Master Thesis.

\appendix

\section{POSITIVITY AND MONOTONICITY OF THE DARK MATTER HALO IN $\gamma$$\gamma$ MODELS}

The condition for the {\it positivity} of the DM halo density profile
$\rhoDM$ in $\gamma \gamma$-models with $0\leq\gamma <3$ is
established from eq. (17) as
\begin{equation}
\MR \ge {(\xi+s)^{4-\gamma}\over \xi (1+s)^{4-\gamma}}, \quad\quad s\ge 0.
\end{equation}
Therefore, $\MR$ must be greater than or equal to the {\it maximum}
$\Rm(\xi,\gamma)$ of the radial function at r.h.s.: note that $\Rm$ is
the {\it minimum} value of $\MR$ in order to have a nowhere negative
DM halo. Simple algebra shows that the maximum is attained at infinity
for $\xi <1$, and at the origin for $\xi >1$, while for $\xi=1$ the
radial function is identically equal to 1.  From eq. (A1) it follows
that
\begin{equation}
\MR \ge \Rm(\xi,\gamma)=\max\left({1\over\xi}, \xi^{3-\gamma}\right), 
\end{equation}
and for $\gamma=2$ we obtain eq. (18).  

The {\it monotonicity} condition for $\rhoDM$ is obtained by requiring
that $d\rhoDM/dr\leq 0$, i.e.
\begin{equation}
\MR\geq {(\xi+s)^{5-\gamma}(\gamma+4s)\over (1+s)^{5-\gamma}\xi(\xi\gamma+4s)},\quad s\ge0.
\end{equation}
Again we must determine the maximum $\Rmon(\xi,\gamma)$ of the r.h.s. of the equation above.
It is easy to show that for $\gamma=0$
\begin{equation}
\Rmon(\xi,0) =\max\left({1\over \xi},\xi^4\right),
\end{equation}
while for $0<\gamma<1$
\begin{equation}
\Rmon(\xi,\gamma) =\max\left({1\over \xi},f_+\right),
\end{equation}
where $f_+$ is the value of the r.h.s. of eq. (A3) at the critical point
\begin{equation}
s_+(\xi,\gamma)= {\sqrt{\gamma[\gamma \xi^2+\xi(5-3\gamma)+\gamma]} -
    \gamma (1+\xi)\over 10}:
\end{equation}
$f_+ \to \xi^4$ and $f_+ \to
\xi^2$ for $\gamma \to 0^+$ and $\gamma \to 1^-$, respectively. Finally
for $1\le \gamma <3$ (and so in particular for JJ models, or for
two-component Hernquist models that could be constructed by using the
same approach of JJ models) it can be shown, quite surprisingly, that
the monotonicity condition coincides with the positivity condition,
and so $\Rmon$ is given by eq. (A2).

The application of the WSC to the isotropic DM halo is obtained from
eq. (26) with $\ra\to\infty$, i.e. $\varrho=\rhoDM$. The condition in absence
of the central BH ($\mu =0$) reduces to  
\begin{equation}
\MR\geq 
{2(s+\xi)^3[6s^3 + 4(1+2\xi)s^2  +(1+7\xi)s +2\xi]\over 
4 \xi (1+s)^4 (3s^2 +3\xi s +\xi^2)}, \quad s\geq 0. 
\end{equation}
For $\xi =1$ the r.h.s. equals 1 independently of $s$. For $\xi\neq 1$
the determination of the maximum leads to study a fifth degree
equation. Fortunately, it can be proved by inspection that the
resulting expression with $s\geq 0$ is negative for $\xi >1$ (and thus
the maximum of eq. [A7] is reached at $s=0$), and positive for $0<\xi
<1$ (and so the maximum is reached for $s\to\infty$). In the two
limits eq. (A7) evaluates to $\xi$ and $1/\xi$, respectively, and so
we conclude that the isotropic DM halo of JJ models (in absence of
central BH) is certainly consistent when $\MR$ satisfies the
positivity and monotonicity condition in eq. (18). We are now in
position to consider the effect of the central BH. A direct analysis
would lead to a cumbersome expression, to be explored
numerically. However, by using the considerations after eq. (30), it
is simple to show that the additional term due to the BH is positive,
and so it reinforces the WSC when the positivity condition on $\rhoDM$
is verified.

The application of the WSC to the OM anisotropic stellar component of
JJ models leads to the study of a seventh degree equation, and shows
that we are in the conditions pertinent to eq. (32). In absence of the
central BH the dependence on $\MR$ disappears,
\begin{equation}
\sa^2\geq -
{s^3 [s^2 +2 (\xi -1)s -\xi]\over 6s^3 +4 (1+2\xi)s^2 + (1+7\xi)s
  +2\xi}, \quad s\geq 0, 
\end{equation}
and we should solve a fifth degree equation when searching for the
maximum of the r.h.s. In Sect. 3.1 we present the results obtained by
numerical inspection of equation above. Restricting further to the
case $\xi =1$ (i.e., reducing to the one-component Jaffe model), the
equation to be solved becomes cubic, with $\sam\simeq 0.1068$ (Ciotti
1999). Finally, the case obtained for $\xi\to 0$ is formally
coincident with the case of a dominant central BH (i.e., only $\Mbh$
is retained in eqs. [29]-[30]), and for this limiting case the WSC
reduces again to a cubic equation, with solution $\sam\simeq 0.31$.

\section{THE LAMBERT-EULER W FUNCTION}

As discussed in Sect.3, for JJ models with central BH it is possible
to invert eq. (24) and express the radius $r$ in terms of the relative
total potential by using the Lambert-Euler $W$ function.  The
integrand in the inversion integral (34) is then obtainedin explicit
and easily tractable form, without resorting to complicate numerical
procedures, because the $W$ function is now fully implemented in the
most used computer algebra systems. The function $W(z)$ (see, e.g.,
Corless et al. 1996) is a multivalued complex function defined
implicitly by the identity
\begin{equation}
W e^{W}=z,
\end{equation}
and the two real branches $W(0,z)$ and $W(-1,z)$ for real values of
$z$ are shown in Fig. B1.
\begin{figure}
\includegraphics[height=0.35\textheight,width=0.45\textwidth]{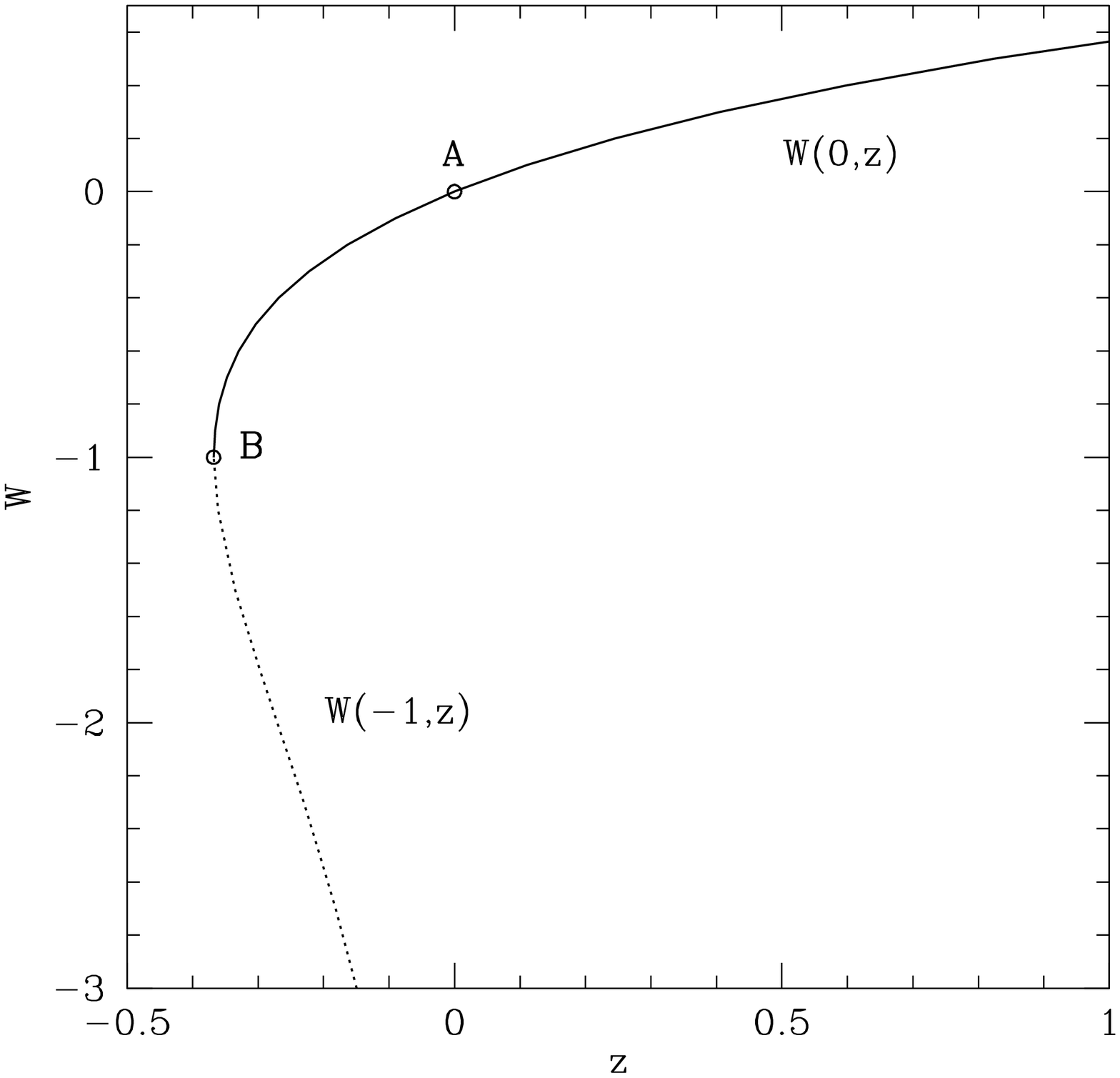}
\caption{The two branches of the real determination of the
  Lambert-Euler $W$ function for real argument.  The coordinates of
  the two marked points are $A=(0,0)$ and $B=(-1/e,-1)$.  The solid
  line represents the branch $W(0,z)$, and only points at right if
  point $A$, with $z\geq 0$, are needed in the inversion formula
  (35). The dotted line is the $W(-1,z)$ branch.}
\label{fig:W}
\end{figure}
With the transformation of variables $W=\left(1+\xi/s\right)\mu/\MR$,
eq. (24) can be rewritten as 
\begin{equation}
W+\ln W = {\xi\psi +\mu\over \MR} +\ln {\mu\over\MR},
\end{equation}
so that eq. (35) is obtained by exponentiation of eq. (B2) and
comparison with eq. (B1). It is immediate to conclude that for the
present problem the relevant branch is given by $W(0,z)$, restricting
to points beyond point $A$.  In fact, when $\psi\to\infty$, the
argument in eq. (B2) is infinite, $W(0,z)\to\infty$, and from eq. (35)
$s\to 0$. Instead, when $\psi\to 0$, the argument tends to $\mu/\MR
e^{\mu/\MR}$, $W \to \mu /\MR$, and so from eq. (35) $s\to\infty$ from
eq. (35), as it should \footnote{By definition of $W$, it follows that
  $W(ye^y)=y$.} .  Finally, note that the derivatives inside the
integral (34) can be expressed analytically in terms of $W$ itself,
because from eq. (B1) it follows that
\begin{equation}
{dW\over dz}={W\over z(1+W)}.
\end{equation}

\section{DF of OM Jaffe model with dominant central BH}

We report the explicit phase-space DF of a Jaffe model with OM
anisotropy and with dominant central BH (i.e., the gravitational field
is produced by the BH only, and the stellar distribution is only a
tracer). The resulting expression can be interpreted as the asymptotic
limit of the DF at high relative energies, i.e. for galactic regions
sufficiently near the central BH. By using the nomenclature in
eq. (34), it is easy to show that the functions $U$ and $V$ can be
written as
\begin{equation}
U(q)={U(\qt)\over\mu^{3/2}},\quad
V(q)={V(\qt)\over\mu^{3/2}},\quad \qt\equiv{q\over\mu},
\end{equation}
where
\begin{equation}
U(\qt)={(16\qt^3+40\qt^2+18\qt+9)\sqrt{\qt}\over 4 (1+q)^3}
-{3(3+8\qt){\rm arcsenh}\sqrt{\qt}\over 4 (1+\qt)^{7/2}},
\end{equation}
and
\begin{equation}
V(\qt)={(13-2\qt)\sqrt{\qt}\over 4 (1+\qt)^3}+
{3(1-4\qt){\rm arcsenh}\sqrt{\qt}\over 4 (1+\qt)^{7/2}}.
\end{equation}
The function $U$ is nowehere negative in the range $0\leq\qt <\infty$,
so the BH dominated Jaffe models are always in the first case
discussed in Sect. 3.1, and only $\sam$ exists: a numerical evaluation
shows that $\sam\simeq 0.082$, in agreement with the trend of the
solid line in Fig. 2 for $\xi\to 0$, when the DM halo ``collapses'' to
a central point mass.

For completeness we also report the explicit DF for the stellar
component of JJ models with $\xi =1$ and in absence of the central
BH, when the resulting expression reduces to the one-component DF in the OM
case. From eq. (34) we now have
\begin{equation}
U(q)={U(\qt)\over\MR^{3/2}},\quad
V(q)={V(\qt)\over\MR^{3/2}},\quad \qt\equiv{q\over\MR},
\end{equation}
with $0\leq\qt <\infty$ and
\begin{equation}
{U(\qt)\over 4\sqrt{2}}=
\Fp (\sqrt{2\qt})+\Fm (\sqrt{2\qt}) -
\sqrt{2}\left[\Fp (\sqrt{\qt}) + \Fm (\sqrt{\qt})\right],
\end{equation}
\begin{equation}
{V(\qt)\over 4\sqrt{2}}=\Fp (\sqrt{2\qt}) - {\Fp (\sqrt{\qt})\over\sqrt{2}},
\end{equation}
where $\Fp (x)=e^{-x^2}\int_0^xe^{t^2}dt$ is the Dawson's function,
and $\Fm (x)=e^{x^2}\int_0^xe^{-t^2}dt=\sqrt{\pi}e^{x^2}{\rm
  Erf}(x)/2$. The functions above, when combined according to
eq. (34), are in perfect agreement with those given by Merritt (1985b,
eq. [6]) and Binney \& Tremaine (2008).  The function $U$ in eq. (C5)
is positive for alla values of $\qt$, as shown by the WSC, but the
function $V$ in eq. (C6) becomes negative for admissible values of
$\qt$, so that $\ra$ cannot be arbitrarily small.  Numerical
evaluation of eq. (32) shows that for consistency $\sa\geq\sam\simeq
0.02205$, in perfect agreement with the solid line in Fig. 2 (obtained
from the general DF) for $\xi =1$. 

From this result one could conclude that the purely radial model does
not exist. However the situation is not so simple. In fact, the DF of
a purely radial model can be written in all generality as
$f=\delta(J^2)h(\Er)$, so that for a finite mass, spatially
untruncated model
\begin{equation}
\rho(\PsiT) ={2\pi\over r^2}\int_0^{\PsiT}{h(\Er) d\Er\over\sqrt{\PsiT
    -\Er}},
\end{equation}
(e.g., Ciotti 2000), and the inversion formula can be immediately
found (e.g., see Richstone \& Tremaine 1984, Oldham \& Evans 2016)
\begin{eqnarray}
h(\Er)&=&{1\over\sqrt{2}\pi^2}{d\over d\Er}
   \int_0^{\Er}
   {\varrho\; d\PsiT\over\sqrt{\Er-\PsiT}}\cr 
   &=&{1\over\sqrt{2}\pi^2}
   \int_0^{\Er}{d\varrho\over d\PsiT}
   {d\PsiT\over\sqrt{Q-\PsiT}},
\end{eqnarray}
where $\varrho =r^2\rho$ is expressed in terms of $\PsiT$, and the
second identity follows from integration by parts when considering
spatially untruncated profiles such those of JJ models.  As shown by
Merritt (1985b, eq. 8) and Evans et al. (2015, eq. 31), for the purely
radial one-component Jaffe model 
\begin{equation}
h(\tilde\Er)= {2\rhon\rs^2\over\pi^2\sqrt{\Psin}}\left[\sqrt{2}\Fp (\sqrt{\tilde\Er}) - \Fp (\sqrt{2\tilde\Er})\right],
\end{equation}
where $\tilde\Er\equiv\Er/\Psin$. The function is positive at {\it
  all} energies, thus showing that the purely radial Jaffe model {\it
  is} consistent. 

These two seamingly contradictory results indicate that the purely
radial case, at least for the Jaffe model, is a singular limit for the
OM parameterization. In practice, we have shown that the non-existence
of the OM (or others) highly radial models cannot by itself exclude
the phase-space consistency of the purely radial configuration.  In
fact, the following argument, built by using the CP92 approach to the
purely radial case, reinforces this conclusion. From the second of
eq. (C8) it follows immediately that a {\it sufficient} condition for
consistency of the purely radial model is that the derivative inside
the integral be non-negative, i.e. in terms of radius
\begin{equation}
{d \varrho (r)\over dr}\leq 0,\quad \varrho(r)=r^2\rho (r).   
\end{equation}
Therefore, in the purely radial model a density profile declining as
$r^{-2}$ or faster at all radii is a {\it sufficient} condition for
consistency (in agreement with the result obtained for the Jaffe
model), while the OM condition (28) (the analogous of eq. (C10) in the
limit of vanishing anisotropy radius) is only {\it necessary} for
phase-space consistency. The mathematical reason of the different
behavior is due to the fact that in eq. (C7) , at variance with the
corresponding expression in the OM case, the preparatory derivative of
the augmented density is not required to perform the Abel inversion.



\end{document}